\documentclass[published]{JHEP3}

\JHEP{11(2003)065}
\received{July 28, 2003}
\revised{October 10, 2003}
\accepted{November 27, 2003}
\keywords{D-branes, Superstring Vacua}

\usepackage{epsfig,psfrag}

\skip\footins = 1\bigskipamount plus 2pt minus 4pt

\def\beq{\begin{equation}}
\def\eeq{\end{equation}}
\def\beqa{\begin{eqnarray}}
\def\eeqa{\end{eqnarray}}
\def\cn{{\cal N}}
\def\IZ{\mathbb{Z}}
\newcommand{\SU}{\mathop{\rm SU}\nolimits}
\newcommand{\SL}{\mathop{\rm SL}\nolimits}
\newcommand{\Tr}{\mathop{\rm Tr}\nolimits}
\renewcommand{\Re}{\mathop{\rm Re}\nolimits}

\title{Saltatory de Sitter string vacua}

\author{Cristina Escoda, Marta G\'omez-Reino and Fernando Quevedo\\
	  DAMTP, Centre for Mathematical Sciences, University of Cambridge\\
          Cambridge CB3 0WA U.K.\\
	  E-mail: \email{C.Escoda@damtp.cam.ac.uk}, \email{M.Gomez-Reino@damtp.cam.ac.uk},
	  \email{f.quevedo@damtp.cam.ac.uk}}

\abstract{We extend a recent scenario of Kachru, Kallosh, Linde and
  Trivedi to fix the string moduli fields by using a combination of
  fluxes and non-perturbative superpotentials, leading to de Sitter
  vacua.  In our scenario the non-perturbative superpotential is taken
  to be either, the racetrack scenario or the ${\cal N}=1^*$
  superpotential for an ${\rm SU}(N)$ theory, originally computed by
  Dorey and recently rederived using the techniques of
  Dijkgraaf-Vafa. The fact that this superpotential includes the full
  instanton contribution gives rise to the existence of a large number
  of minima, increasing with $N$.  In the absence of supersymmetry
  breaking these correspond to supersymmetric anti de Sitter vacua.
  The introduction of antibranes lifts the minima to a chain of
  (non-supersymmetric) de Sitter minima with the value of the
  cosmological constant decreasing with increasing compactification
  scale. Surprisingly a similar picture occurs for the simpler system
  of the racetrack scenario. The relative semiclassical stability of
  these vacua is studied. Possible cosmological implications of these
  potentials are also discussed.}

\begin{document}

\section{Introduction}\label{section1}

Supersymmetry breaking and moduli fixing have been the main obstacles
for string theory to make contact with low energy physics. These
questions are also essential for the study of any possible
cosmological implications of the theory.  Over the years there have
been several proposals to solve these problems. Supersymmetry breaking
mechanisms include the effect of non-perturbative field theoretical
effects such as gaugino condensation and, more recently, the explicit
breaking\pagebreak[3] due to the presence of antibranes or brane intersections in
low scale string models.  The remnant potential for the moduli fields
in these cases is such that the global minimum is either anti de
Sitter space with a very large (negative) vacuum energy or the
potential is of the runaway type, towards infinite extra dimension
and/or zero string coupling.  The problem is then not how to break
supersymmetry but actually what to do, after breaking it, with the
remaining potential for the moduli.

Recently there has been interesting progress towards the solution of
the continuum vacuum degeneracy problem in string theory.  The
introduction of fluxes of RR or NS-NS
forms~\cite{fluxes}--\cite{kst} has proven to be very efficient to
fix many of the moduli fields, including the dilaton. However in
typical orientifold (or $F$-theory) compactifications, the overall
K\"ahler modulus cannot be fixed by this effect~\cite{drs,gkp}.

An interesting proposal to also fix this modulus, due to Kachru,
Kallosh, Linde and Trivedi (KKLT)~\cite{kklt}, combines the fluxes
with non-perturbative superpotentials that have been discussed in
string theory in the past, consisting of a single exponential of the
corresponding modulus.  The modulus is then stabilised for a
supersymmetric anti de Sitter point.  The further inclusion of anti D3
branes breaks supersymmetry explicitly and can lift the minimum of the
potential to a de Sitter minimum with varying value of the
cosmological constant, depending on the different parameters of the
theory.  Even though this is a very simple set-up, involving much fine
tuning and the simplest non-perturbative superpotential, it represents
a concrete step forward towards fixing the moduli after supersymmetry
breaking, with a potentially realistic value of the vacuum energy.
 
In a different direction, progress has also been made in the
understanding of non-perturbative effects in supersymmetric field
theories. In particular, several techniques have been developed that
allow the computation of the exact non-perturbative
superpotentials. In some simple cases they have been derived in closed
form encoding the infinite sum of exponentials of the inverse gauge
coupling expected from the full instanton and fractional instanton
effects.
 
In this note, we slightly extend the KKLT proposal by considering more
general non-perturbative superpotentials than the single exponential
considered in KKLT.  In particular we consider the non-perturbative
superpotential for a supersymmetric theory for which the exact
superpotential has been computed explicitly, including the infinite
instanton sum. This is the so-called ${\cal N}=1^*$
theory~\cite{dorey,ps}.

\looseness=1The structure of the potentials generated in this way is such that
there are many AdS supersymmetric minima in the absence of the
antibranes and many de Sitter minima in their presence, with
intermediate configurations having both dS and AdS vacua, all
non-supersymmetric.  The minima are such that the value of the vacuum
energy decreases with increasing value of the compactification scale.
This rich vacuum structure may have interesting physical implications
and can serve as a concrete nontrivial example illustrating the
possible `landscape' of string theory~\cite{susskind}.  It is also
similar to the staircase potentials proposed by Abbott~\cite{abbott}
and to the models presented in~\cite{tb,bp,salta} exhibiting a
dynamical relaxation of the cosmological constant. Furthermore, there
is enough freedom to fine tune one of the minima to have a
cosmological constant as small as required. The presence of several de
Sitter minima could also lead to interesting realisations of
inflation.

We also consider simpler cases in which the superpotential is a finite
sum of exponentials as it appears in the much studied racetrack
scenarios~\cite{krasnikov}. These are the simplest extensions of the
mechanism of KKLT and, in the cases that lead to several minima, they
serve as simple examples to study the transition between different
vacua.

This article is organised as follows: After briefly reviewing the
effect of fluxes to fix the moduli in the next section, we start
considering in section~\ref{section3}, the general scalar potential
for the K\"ahler modulus field for arbitrary superpotential. In
section~\ref{section4} we first consider the case of superpotentials
with two exponentials which can give rise to one or two different
minima. Section~\ref{section5} discusses the potential for the $\cn
=1^*$ theory with its rich vacuum structure with and without breaking
supersymmetry.  Section~\ref{section6} is dedicated to the stability
analysis of the different minima. We recover in particular the results
of KKLT about the life time of the dS minimum with smallest positive
value of the cosmological constant which is much larger than the age
of the universe but smaller than the Poincar\'e recurrence time. We
also discuss the relative probability for tunneling between different
minima.
 
\section{Fluxes and moduli fixing}\label{section2}

Type-IIB strings have RR and NS-NS antisymmetric 3-form field
strengths, $H_3$ and $F_3$ respectively,
that can have  a (quantised) flux on 3-cycles of the  compactification
manifold.
\beqa
\frac{1}{4\pi^2 \alpha'} \int_A F_3 & = & M \,,
\nonumber\\
\frac{1}{4\pi^2 \alpha'} \int_B H_3 & = & -K\,,
\eeqa
where $K$ and $M$ arbitrary integers and $A$ and $B$ different
3-cycles of the Calabi-Yau manifold.

The inclusion of fluxes of RR and/or NS-NS forms in the compact space
allows for the existence of warped metrics that can be computed in
regions close to a conifold singularity of the Calabi-Yau manifold,
with a warp factor that is exponentially suppressed, depending on the
fluxes, as:
\beq
a_0\sim e^{-2\pi K/3g_s M} \, .
\eeq
Therefore fluxes can naturally generate a large hierarchy.  Here $g_s$
is the string coupling constant.

Fluxes have also proven very efficient for fixing many of the string
moduli, including the axion-dilaton field of type-IIB theory $S=
e^\phi +i{\hat a}$.  A very general analysis of orientifold models of
type IIB, or its equivalent realisation in terms of $F$-theory, has
been done by Kachru and collaborators~\cite{gkp,kst}. In the $F$
theory approach, the geometrical picture corresponds to an
elliptically fibered four-fold Calabi-Yau space $Z$ with base space
${\cal M}$ and the elliptic fiber corresponding to the axion-dilaton
field $S$.

The consistency condition in terms of tadpole cancellation implies a
relationship between the charges of D-branes, O-planes and fluxes that
can be written as follows:
\beq
\label{tadpole}
N_{{\rm D}3}-N_{\bar {\rm D}3}+ N_{\rm flux}= \frac{\chi(Z)}{24}\,,
\eeq
where the left hand side counts the number of D3 branes and antibranes
as well as the flux contribution to the RR charge:
\beq
N_{\rm flux}= \frac{1}{2\kappa_{10}^2 T_3} \int_M H_3\wedge F_3\,.
\eeq
The r.h.s.\ of~(\ref{tadpole}) refers to the Euler number of the
four-fold manifold $Z$ or in terms of orientifolds of type IIB, to the
contribution of the D3-brane charge due to orientifold planes and
D7-branes.  Here $\kappa_{10}$ refers to the string scale in 10D and
$T_3$ to the tension of the D3 branes.

The fluxes generate a superpotential on the low-energy
four-dimensional effective action of the Gukov-Vafa-Witten
form~\cite{gvw}:
\beq
W = \int_M G_3\wedge \Omega\,,
\eeq
where $G_3=F_3-iS H_3$ with $S$ the dilaton field
and $\Omega$ is the unique $(3,0)$ form of  the
corresponding Calabi-Yau space.

In the simplest models there will be one single K\"ahler structure
modulus defining the overall size of the Calabi-Yau space which we
denote by $T=r^4+ib$ where $r$ is the scale of the extra dimensions
and $b$ an axion field coming from the RR 4-form
($T=i\rho$ in the conventions of~\cite{gkp,kklt}). The relevance of
this modulus is that it is the one that cannot be fixed by the
fluxes. Its K\"ahler potential is of the no-scale form, that is:
\beq
K = \tilde K (\varphi_i, \varphi_i^*) -3 \log \left( T+ T^*\right),
\eeq
with $\tilde K$ the K\"ahler potential for all the other fields
$\varphi_i$ except for $T$. This implies  that the supersymmetric 
scalar potential takes the form
\beq
V_{\rm SUSY} = e^K\left(K^{i\bar j} D_iW \bar{D_{j}}{\bar W}\right),
\eeq
with $K^{i\bar j}$ the inverse of the K\"ahler metric $K_{i \bar j}=
\partial_i \partial_{\bar j} K$ and $D_iW =\partial_i W + W\partial_i
K$ the K\"ahler covariant derivative. The $T$ dependence of the
K\"ahler potential is such that the contribution of $T$ to the scalar
potential cancels precisely the term $-3e^K|W|^2$ of the standard
supergravity potential, this is the special property of no-scale
models~\cite{noscale}. Since this potential is positive definite, the
minimum lies at zero, with all the fields except for $T$ fixed from
the conditions $D_iW=0$. This minimum is supersymmetric if $D_TW=W=0$
and not supersymmetric otherwise.

Since the superpotential does not depend on $T$, we can see in this
way that the fluxes can fix all moduli but $T$. In order to fix $T$
KKLT procede as follows:

\begin{enumerate}
\item Choose a vacuum in which supersymmetry would be broken by the $T$
  field, such that $W=W_0\neq 0$.

\item Consider a non-perturbative superpotential generated by
  euclidean D3-brane or by gaugino condensation due to a non-abelian
  sector of wrapped $N$ D7-branes. For which the gauge coupling is
  $\frac{8\pi^2}{g_{\rm YM}^2}= 2\pi \frac{r^4}{g_s} = 2\pi\Re T$.
  Which induces a superpotential of the form $W_{np}= B e^{-2\pi
    T/N}$.  Combining the two sources of superpotentials $W_0 +
  W_{np}$, they find an effective scalar potential with a non-trivial
  minimum at finite $T$ and the standard runaway behaviour towards
  infinity, as usual. The non-trivial minimum corresponds to negative
  cosmological constant giving rise to a supersymmetric AdS vacuum.

\item In order to obtain de Sitter vacua, KKLT, consider the effect of
  including anti D3 branes, still satisfying the
  condition~(\ref{tadpole}).  This has the net effect of adding an
  extra (non-supersymmetric) term to the scalar potential of the form:
\beq
V  = V_{\rm SUSY} + \frac{D}{(\Re T)^3}
\label{brsusy}
\eeq

with the constant $D=2 a_0^4 T_3/g_s^4$ paramterizing the lack of
supersymmetry of the potential. Here $a_0$ is the warp factor at the
location of the anti D3 branes and $T_3$ the antibrane tension. The
net effect of this is that for suitable values of $D$ the original AdS
minimum gets lifted to a dS one with broken supersymmetry. See
figure~\ref{fig1}.
\end{enumerate}

\EPSFIGURE[r]{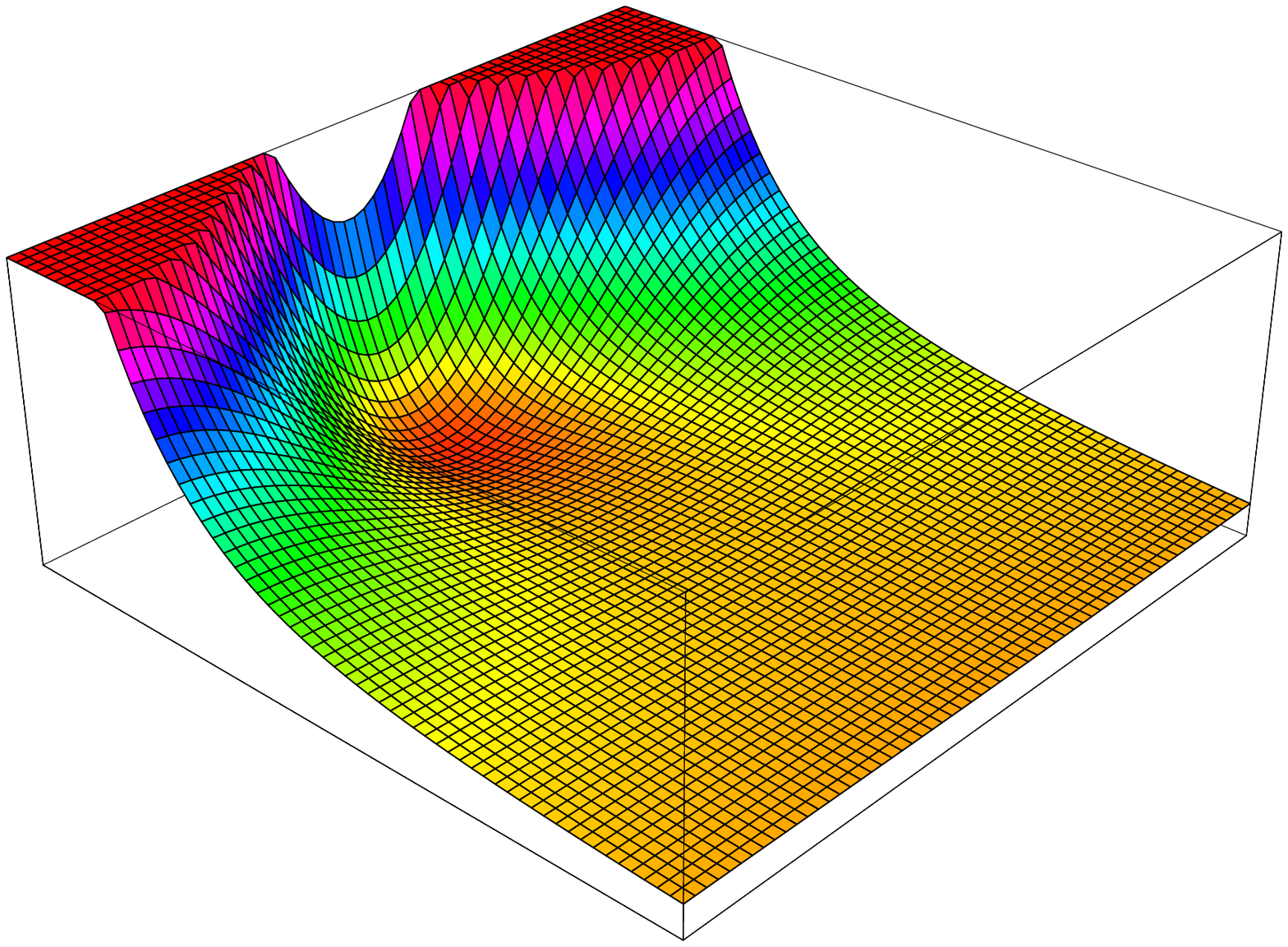,width=8cm}{The scalar potential considered
  in~\cite{kklt} with one single de Sitter minimum.\label{fig1}}

{\sloppy Here we will modify the KKLT scenario in two ways. First, regarding
the original fluxes, we can consider the supersymmetric configuration
where $W_0=0$ with the form $G_3$ being of the $(2,1)$ type. That is
we may include a non-vanishing $W_0$ part in the superpotential but it
will not be necessary.  Second, regarding the non-perturbative
superpotential, we explore the simplest ${\cal N}=1$ supersymmetric
model for which the full non-perturbative superpotential has been
computed, including the contribution from all (infinite)
instantons. This is the so-called ${\cal N}=1^*$ model, constructed
from mass deformation of ${\cal N}=4$ super Yang-Mills.

}

Our first minor modification allows to start with an explicit
supersymmetric model, before considering the low-energy
non-perturbative effects. This avoids the need to fine tune the value
of $W_0$ in looking for non-trivial minima.  Our second modification
allows exploring the possibility of an exact instanton calculation,
instead of a single instanton calculation as it is usually
considered. We will see that this exact superpotential will have a
constant piece (similar to $W_0$) and an infinite sum of exponential
terms, allowing for a very rich vacuum structure.  For completeness,
we also considered simpler superpotentials including a sum of two
exponentials as in the racetrack scenario.

\section{The general scalar potential}\label{section3}

\subsection{The supersymmetric potential}\label{section3.1}

The standard $\cn=1$ supergravity formula for the potential in Planck
units reads
\beq
V_{\rm SUSY} = e^{K}\left( \sum_{i,j} K^{i\bar j} D_{i}W \overline {D_j W}
- 3|W|^2 \right),
\label{treepot}
\eeq
where $i,j$ runs over all moduli fields. As we already mentioned,
$K^{i\bar j}=\partial_i\partial_{\bar j}K$ where $K$ is the
corresponding K\"ahler potential and
$D_iW=\partial_iW+(\partial_iK)W$.  In our case we are working with a
model having only one K\"ahler modulus, (that is, $h^{1,1}(M) = 1$) as
we will be focusing on the $T$ field. All other fields are assumed to
have been fixed by the fluxes just as in~\cite{kklt} so the
superpotential $W$ will depend on the superfield $T$.

Then our purpose is to study the scalar potential $V(T)$ which also
depends on the K\"ahler potential.  We take the weak coupling result
in 4-dimensional string models, namely
\beq
K=-3\log (T+T^*)\,, 
\label{kpdef}
\eeq
and neglect possible perturbative and non-perturbative corrections.
For simplicity of notation we will write the field $T$ in terms of
its real and imaginary parts:
\beq
T\, \equiv X+i Y\,.
\eeq

Using~(\ref{treepot}) and~(\ref{kpdef}) the scalar potential turns out to be
\beq
V_{\rm SUSY} = \frac1{8 X^3} \left\{ \frac13|2X W^\prime - 3 W|^2 -
3|W|^2 \right \},
\label{spot}
\eeq
where by $^\prime$ we understand derivatives with respect to the field
$T$. To compute the supersymmetric minima of the scalar potential we
need to calculate the derivative of~(\ref{spot}) that is given by
\beq
V^\prime=\frac{\partial V}{\partial T} = \frac{ (2X
W^{\prime\prime} -2W^\prime)(2X W^\prime - 3 W)^* - 2(W^\prime)^* 
(2X W^\prime - 3W)}{24 X^3}\,,
\label{vloc}
\eeq

Notice then that there can be two types of extrema. The supersymmetric
extrema appear when
\beq
2X W^\prime - 3 W = 0\,.
\label{msloc}
\eeq
In this case the value of the potential at the extremum is clearly
negative definite leading to an anti de Sitter vacuum.  These extrema
are minima provided that:
\beq
|X W^{\prime\prime} -W^\prime| >  |W^\prime|\,.
\eeq

The nonsupersymmetric extrema occur at
\beq 
(X W^{\prime\prime} -W^\prime) =  (W^\prime)^* e^{2i\gamma}\,, 
\label{mnsloc}
\eeq
where $\gamma=\arg(2X W^\prime - 3W)$. 

In all cases considered here, our analysis shows that the
condition~(\ref{mnsloc}) is never fulfilled at the minima. Then all
minima are supersymmetric and lead to negative cosmological constants.

\subsection{Supersymmetry breaking}\label{section3.2}

Following the lines of~\cite{kklt}, to uplift the anti de Sitter vacua
to de Sitter vacua we will introduce in our model a ${\overline {{\rm D}3}}$. This
will break the supersymmetry of the system. The introduction of the
antibrane does not introduce extra translational moduli as its
position is fixed by the fluxes~\cite{kpv}, so it just contributes to
the energy density of the system. This contribution is given by~\cite{kpv}
\beq
\delta V = { D\over X^3}\label{sb}\,,
\eeq
where the coefficient $D$ is a function of the tension of the brane
$T_3$ and of the warp factor $a_0$ and has the form 
$D=2\frac{a_0^4T_3}{g_s^4}$, where $g_s$ denotes the string
coupling. The coefficient $D$ depends on the fluxes through the
dependence of the warp factor $a_0$ on them and it is therefore
quantised, although the range of values of the fluxes can be such that for
practical purposes it may be considered as an almost continuum
variation~\cite{kklt,frey}. 

If we add this supersymmetry breaking term to the scalar
potential~(\ref{spot}) we have now that the expression for the scalar
potential is given by
\beq
V = V_{\rm SUSY} + \delta V\,.
\label{potproto}
\eeq
The effect of the supersymmetry breaking term depends on the range of
values of the parameter $D$. If $D$ is very small, the potential will
not change substantially and the minima remain anti de Sitter. For a
critical value of $D$ one of the minima will move up to zero vacuum
energy and then to de Sitter space. Continuing increasing $D$, more
minima become de Sitter until all of them are de Sitter. If $D$ is
larger than another critical value the nonsupersymmetric term starts
dominating the potential and starts eliminating the different extrema
to make the potential runaway with $X$. The range of values of $D$
will depend very much on the form of $W$ and we will illustrate the
behaviour in the examples below.

\section{Superpotentials with two exponentials}\label{section4}

Our main interest in this paper is to consider non trivial
superpotentials including all instanton contributions in terms of an
infinite sum of exponentials. For simplicity we will first discuss the
simpler case of superpotentials with just two exponentials. We then
consider the following superpotential
\beq
\label{dosexp}
W=W_0+A\,e^{-aT}+B\,e^{-bT}\,.
\eeq

\subsection{The racetrack scenario}\label{section4.1}

This superpotential has an interest by itself since it includes the
standard racetrack scenario (when $W_0=0$), very much discussed in
order to fix the dilaton field at weak coupling.  The origin of the
so-called racetrack scenario is the condensation of gauginos of a
product of gauge groups. For an $\SU(N)\times \SU(M)$ group we would
have $a=2\pi/N$ and $b=2\pi /M$. For large values of $N$ and $M$, with
$M$ close to $N$ the minimum is guaranteed to lie in the large field
region. A similar result will happen for the field $T$ taken to be the
gauge coupling at a product gauge group coming from the D7 sector.
Turning on $W_0$ it is possible to find more than one minimum that,
following the KKLT method, would then lead to many (order 10) de
Sitter vacua depending on the values of the parameters $W_0, N, M, A,
B$. In figure~\ref{fig2} we present a contour plot of the potential
illustrating this behaviour with $W_0\sim - 10^{-3}$ and all other
constants in the range $1,10$.

\subsection{A finite instanton sum}\label{section4.2}

The potential takes a  manageable form if we assume $b=2a$ which
allows more than one minima and we will follow  mostly for 
illustration purposes. In this case
using~(\ref{treepot}),~(\ref{kpdef}) and~(\ref{dosexp}) the scalar 
potential turns out to be
\beqa
V_{\rm SUSY}  &=&  \frac{ae^{-4aX}}{6 X^2} \Bigl[  6 B\,e^{2a X}W_0\cos
(2aY)+6B^2+3A^2\,e^{2aX}+4aB^2X+
\nonumber\\&&
           \hphantom{\frac{ae^{-4aX}}{6 X^2} \Bigl[}\!
+aA^2 X\,e^{2a X}+A\,e^{a X}(3\,e^{2aX}W_0+B(9+4aX))\cos(aY) \Bigr].
\label{potdosexp}
\eeqa
Notice that this potential is periodic in $Y$ with period $2\pi/a$,
and is also invariant under the reflexion $Y\rightarrow -Y$.

\DOUBLEFIGURE[t]{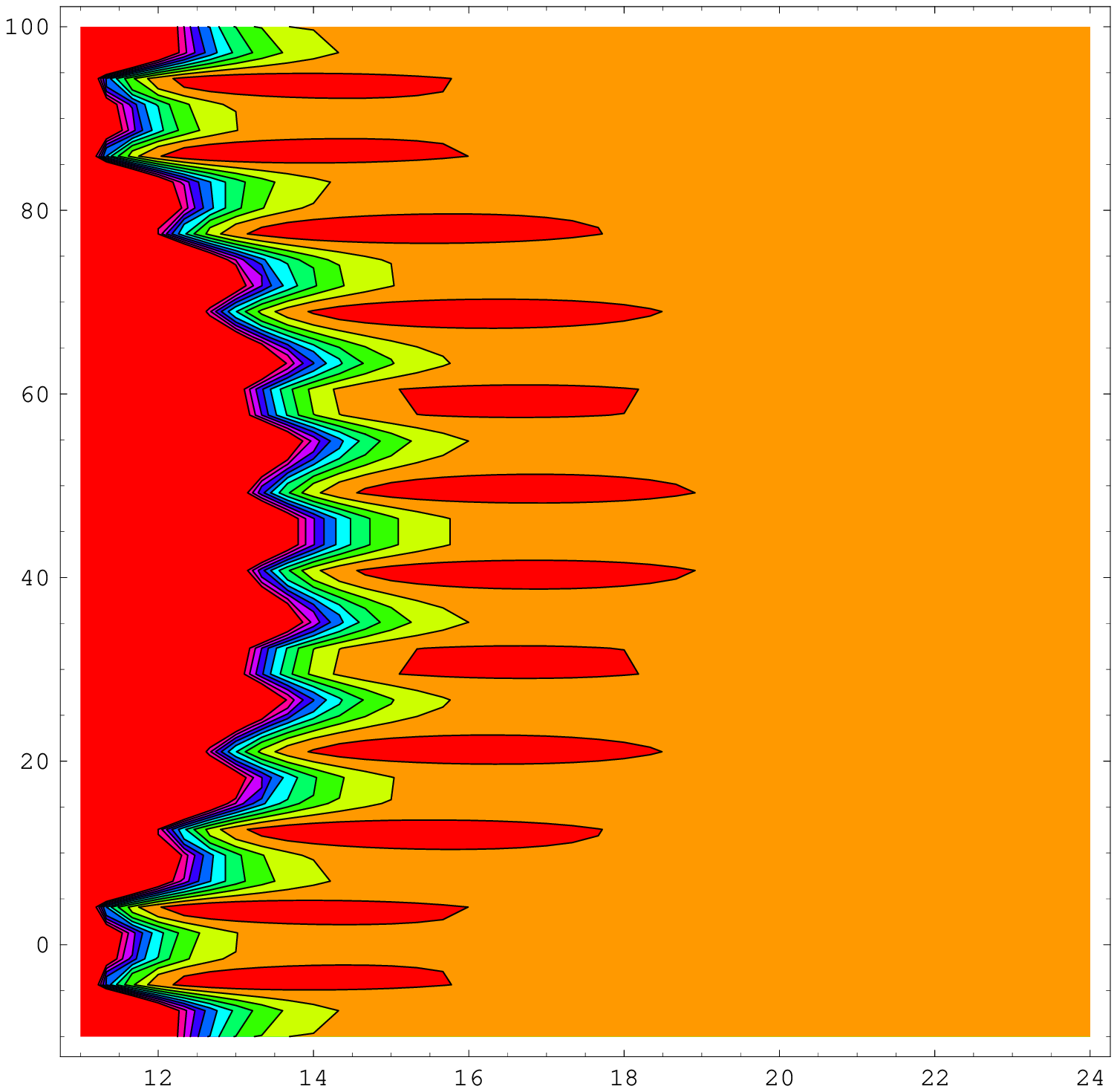,width=.48\textwidth}
{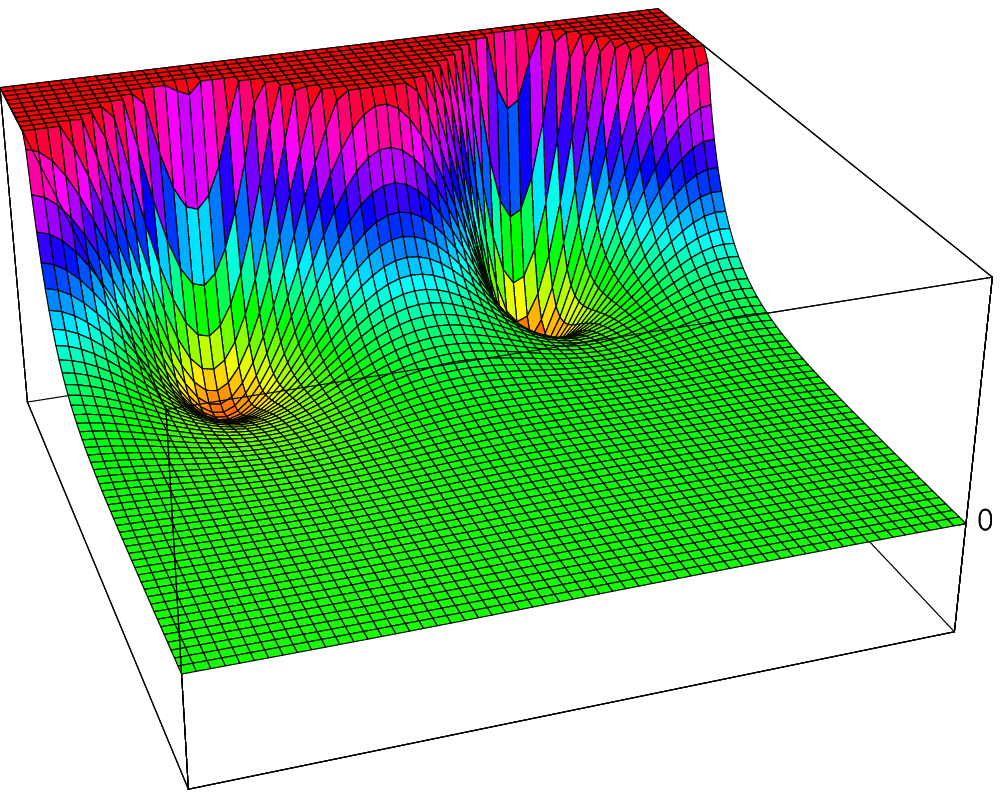,width=.48\textwidth}
{Contour plot for a racetrack type potential with several
  minima.\label{fig2}}
{Graph of the scalar potential with $a=0.1$, $A=1$, $B=5\cdot 10^4$
  and $W_0=-10^{-3}$.\label{fig3}}

From~(\ref{potdosexp}) it is easy to see that if $W_0$ is negative
($W_0<0$) the minima of the potential will be located at
$Y=Y^{(n)}=\pi n/a$, $n=0,\pm1,\pm2,\ldots $ For each value of
$Y^{(n)}$ we will find a minimum at $X=X^{(n)}$.  Since the
potential~(\ref{potdosexp}) is periodic in $Y$ with period $2\pi/a$ we
will just consider the cases $n=0$ and $n=1$, being the rest of the
cases copies of these two. We will denote these minima as
$(X^{(\pm)},Y^{(\pm)})$, where $+$ will denote the case $n=0$ and $-$
will denote the case $n=1$. In figure~\ref{fig3} we show an explicit
example of the potential~(\ref{potdosexp}).

At the minima of the potential we have that
\beq
\partial_{X}V\mid_{Y=Y^{(\pm)}}=0\quad\Longrightarrow\quad
W_0=-\frac{e^{-2aX^{(\pm)}}}{3}\left( \pm A\,e^{aX^{(\pm)}}
(3+2aX^{(\pm)})+B(3+4aX^{(\pm)})\right).
\label{Om0}
\eeq 
Using this we can get the values of the potential at the minima. Those
values are given by
\beq
V_{0}^{(\pm)}=-\frac{a^2}{6X^{(\pm)}}\,
(\pm A\,e^{-aX^{(\pm)}}+2 B\,e^{-2aX^{(\pm)}})^2\,.
\label{uno}
\eeq
Note that from~(\ref{Om0}) we find that $X^{(+)}>X^{(-)}$ and then
from~(\ref{uno}) $V_{0}^{(+)}>V_{0}^{(-)}$. Also note from~(\ref{uno})
that we will always have two non-degenerate minima with null or
negative value of the potential, that is, we will have either
Minkowski or AdS vacua.

On the other hand, if $W_0$ is positive ($W_0>0$), the points located
at $Y=Y^{(n)}=\pi n/a$, $n=0,\pm1,\pm2,\ldots $ are no longer minima,
but maxima. In this case we find that the scalar potential will have
two degenerate minima in every $2\pi/a$ period of $Y$. Those minima
fulfill $Y^{(1)}+Y^{(2)}=2\pi/a$ and $X^{(1)}=X^{(2)}$ (the reason for
that is the invariance of~(\ref{potdosexp}) under $Y\rightarrow
-Y$). This means that we should need more exponential terms
in~(\ref{potdosexp}) in order to have at least two non-degenerate
minima. Therefore we will concentrate in the case $W_0<0$, as is the
simplest case of that type. The analysis of the case $W_0>0$ is
analogous to the case considered in~\cite{kklt}, where they analyse
the case of a scalar potential with just one minimum.

Now, we will study what is the effect of breaking supersymmetry in
this model. The introduction of the supersymmetry breaking
term~(\ref{sb}) acts in the following way: when $D/(X^{(\pm)})^3 \ll
V_{0}^{(\pm)}$ the potential is almost not affected, and we still have
two AdS minima.  If we increase the value of $D$ such that
$D/(X^{(\pm)})^3\sim V_{0}^{(\pm)}$, the value of the scalar potential
at the minima increases with $D$, so the minima will eventually become
positive (dS minima). For larger values of $D$ the minima become
saddle points and then disappear.

Also notice that the supersymmetry breaking term~(\ref{sb}) does not
involve $Y$, therefore the new potential will also be periodic in $Y$
with the same period as before, and its minima will also be located at
$Y=Y^{(n)}=\pi n/a$, $n=0,\pm1,\pm2,\ldots $ Again, for each value of
$Y^{(n)}$ we will find a minimum in $X$ but now at different values
than in the supesymmetric case. We will denote these values as ${\hat
  X}^{(n)}$. For each of those values we will find again two different
values of the potential. It is interesting to note that the position
of the minima in the non-supersymmetric case ($X={\hat X}^{(\pm)},
Y=Y^{(\pm)}$) does not vary significantly from the position of the
minima in the supersymmetric case ($X={X}^{(\pm)}, Y= Y^{(\pm)}$).

Therefore, for a given value of $D$ such that $D/(X^{(\pm)})^3\sim
V_{0}^{(\pm)}$ the non-susy scalar potential will have two different
minima that, depending on the value of the constants, can be either
positive or negative. If we want to describe the present stage of the
acceleration of the universe within this framework, we would like both
minima to be dS minima, such that $V_0^{(+)}>V_0^{(-)} \sim 10^{-120}$
in Planck units. This can always be done in this model for example by
fine-tuning the value of $D$. In figure~\ref{fig4} we show an explicit
example where the two AdS minima that appear in the supersymmetric
case shown in figure~\ref{fig3} become two dS minima in the
non-supersymmetric case.

\FIGURE[t]{\begin{tabular}{cc}
\epsfig{file=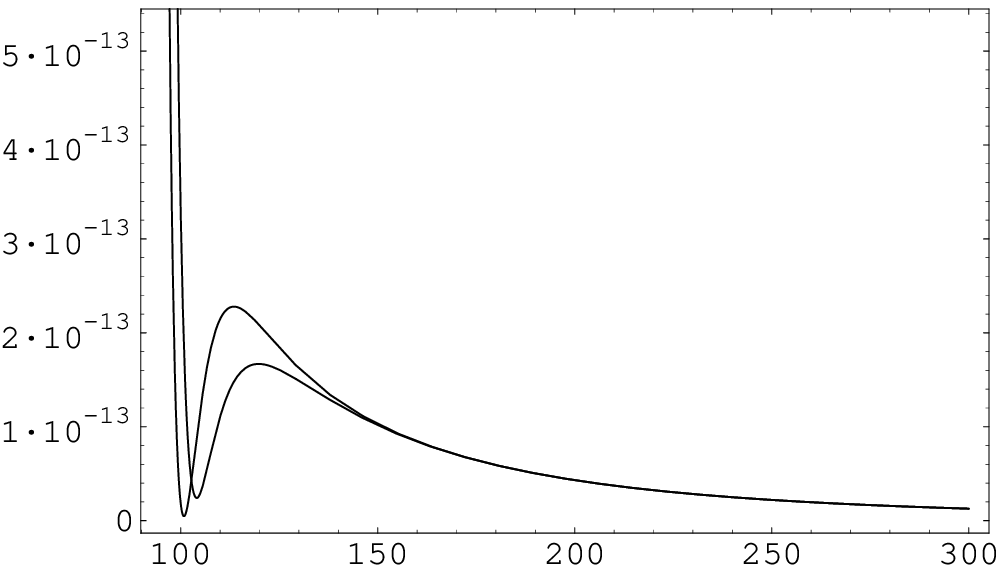, width=.47\textwidth}
&
\epsfig{file=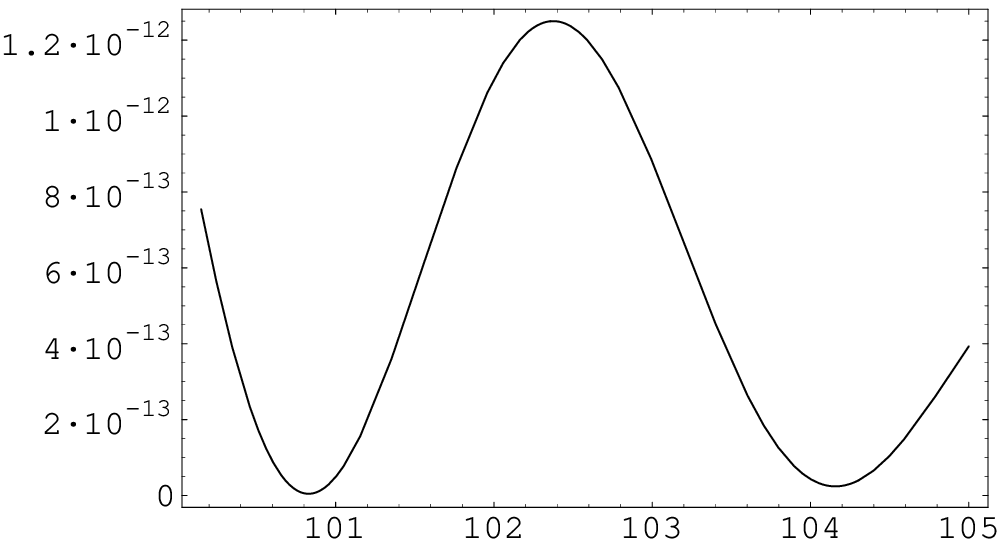, width=.47\textwidth}
\\
\parbox{.47\textwidth}
{\small($a$) Cross section of the scalar potential at $Y=0$ and 
$Y=\pi/a$.}
&
\parbox{.47\textwidth}
{\small($b$) Cross section of the scalar potential along a line that joins
  the two minima.}
\end{tabular}%
\caption{Example of a configuration with $a=0.1$, $A=1$, 
$B=5\cdot 10^4$, $W_0=-10^{-3}$, and $D=3.5\cdot 10^{-7}$. In this case we 
find a configuration with two de Sitter minima.\label{fig4}}}

\section{$\cn=1^*$ theory}\label{section5}

In this section, we would like to study more general superpotentials
were the full instanton sum has been computed, being the simplest such
case the $\cn=1^*$ theory. The $\cn=1^*$ theory is mass deformed
$\cn=4$ super Yang-Mills in which the $\cn=1$ chiral multiplets inside
the $\cn=4$ Yang-Mills multiplet are given a mass. At the moment we do
not have a concrete example where this theory appears at low energies
on the D7-branes.  We may think of ways that it could arise, for
instance since the D-branes break one half of the supersymmetries,
they tend to have a $\cn =4$ gauge theory inside. Having a model that
breaks supersymmetry to $\cn=1$ outside the brane would naturally
induce masses to the chiral multiplets. However here we will take the
$\cn=1^*$ superpotential only as an illustrative example of what we
can expect from full instanton contributions to the superpotential.

\subsection{Background}\label{section5.1}

Let us first review what the $\cn=1^*$ theory is, starting from
$\SU(N)$, $\cn=4$ super Yang-Mills. The $\SU(N)$, $\cn=4$ super
Yang-Mills theory can be written in terms of $\cn=1$ superfields as a
gauge theory with three massless chiral superfields $\Phi_i$ in the
adjoint of $\SU(N)$ with superpotential
\beq
W = \epsilon_{ijk} \Tr \Phi_i \left[ \Phi_j, \Phi_k\right].
\eeq
A deformation of this theory by adding mass terms to these superfields
\beq
\Delta W = m_i \Tr \Phi_i^2\,,
\eeq
with all $m_i\neq 0$ breaks supersymmetry to $\cn=1$. This is the
$\cn=1^*$ theory. There are further deformations of the original $\cn
=4$ theory that may be considered~\cite{kumar2}.

The classical vacua of this theory can be found by solving $\partial
W_T/\partial \Phi_i=0$ for $W_T=W+\Delta W$, which leads to 
\beq
\left[\Phi_i, \Phi_j\right] = \epsilon_{ijk} m_k \Phi_k
\eeq
and therefore the solutions correspond to $N$-dimensional
representations of the $\SU(2)$ algebra, giving rise to the different
phases of the theory.

The massive (Higgs and confining) phases of this theory are well
understood. They are labelled by a triplet of integers $(p,q,k)$ with
$pq=N$ and $0<k<q$. These phases interpolate from the full confining
phase $q=N$ to the full Higgs phase $p=N$.  The exact superpotential
for $\cn=1^*$ was derived by using instanton techniques for the theory
compactified on a circle. The compactification to three dimensions is
a computational trick and it turns out that the superpotential is
independent of the compactification radius~\cite{dorey}.  After
integrating out the gauge fields, the superpotential depends only on
the (complex) gauge coupling $\tau$ and takes the form:\footnote{An
  overall factor of order $m^3=\frac{N^3}{24} m_1 m_2m_3$ appears
  multiplying the superpotential, where $m_i$ are the masses of the
  chiral superfields of the $\cn=4$ theory.  This will scale the
  scalar potential by a particular scale, which we take as unity for
  simplicity, but need to keep it in mind when combining with the
  non-supersymmetric case and to take care of actual number estimates
  regarding the value of the cosmological constant. For consistency we
  have to take the scale set by the $m_i$'s to be hierarchically
  smaller than the string scale. It is not yet clear if these small
  mass parameters will be achieved in explicit models.}
\beq
\label{master}
W_{p,q,k}(\tau)= E_2(\tau) - \frac{p}{q}
E_2\left(\frac{p}{q}\tau+\frac{k}{q}\right).
\eeq
Here and later in this section by $E_n$ we will denote the Eisenstein
modular functions.

The modular properties of the superpotential show that
under a $\SL(2,\IZ)$ transformation 
\beq
\tau \rightarrow \frac{a\tau +b}{c\tau +d}
\eeq
the effective theory in a given phase is not invariant but it
exchanges different phases by changing the values of $p,q,k$. For
instance $\tau\rightarrow \tau+1$ requires $(p,q,k)\rightarrow (p,q,
k+p)$ and $\tau\rightarrow -1/\tau$ implies $(p,q,k)\rightarrow
(\alpha, \frac{N}{\alpha},k')$ with $\alpha =\gcd(q,k)$ and $k'$ having
a complicated dependence on $p,q,k$ with $k'=0$ if $k=0$, indicating
the exchange of Higgs and confining phases under this transformation.

It is interesting to note that the validity of this exact
non-perturbative result has been checked using string theory
techniques~\cite{ps} and more recently from Dijkgraaf-Vafa
techniques~\cite{dv,kumar2}, making this expression very robust.

\subsection{The scalar potential}\label{section5.2}

Following~\cite{fkq} we will now promote the parameter $\tau$ to a
full $\cn=1$ superfield that we identify with the modulus
field\footnote{In reference~\cite{fkq} the corresponding field was the
  dilaton $S$ rather than $T$ then the difference between the
  potential calculated there and the supersymmetric potential we
  consider here is only in the factor of $3$ in the K\"ahler
  potential. The main difference appears when we consider the
  nonsupersymmetric case.} $T=-i\tau$. Then the
superpotential~(\ref{master}) can be written in terms of $T$ as
\beq
W_{p,q,k}(T)=E_2(T) - \frac{p}{q} E_2\left(\frac{p}{q} T-i\frac{k}{q}\right),
\label{wdef}
\eeq

The superpotential $W_{p,q,k}(T)$ transforms as a modular form
of weight $2$ once the values of $p,q,k$ are transformed
accordingly. As we already mentioned, the net effect of a $\SL(2,\IZ)$ 
transformation is that it  changes one massive phase to a different phase.

For concreteness we will concentrate in the confining phase $p=1$, 
$q=N$, for which it is enough to set $k=0$, since $k\neq0$ can be
reached by a translation $T\rightarrow T-ik$. In this case the scalar
potential will be given by~(\ref{spot}) where now the superpotential
and its derivative take the form
\beqa
W(T)&=&E_2(T) - \frac{1}{N} E_2\left(\frac{T}{N}\right),
\label{w8def}\\
W'(T)&=&\frac{\pi}{6}\left\{(E_4(T)-E_2(T)^2)-\frac{1}{N^2}
\left(E_4\left(\frac{T}{N}\right)-E_2\left(\frac{T}{N}\right)^2\right)\right\}.
\eeqa
We will mostly work with the $E_n(T)$ expansions in terms of the
variable $q=e^{-2\pi T}$ that are given by
\beq
E_2 = 1 - 24 \sum_{n=1}^\infty \sigma_1(n) q^n\,,  
\qquad
E_4 = 1 + 240 \sum_{n=1}^\infty \sigma_3(n) q^n\,, 
\label{efns}
\eeq
where $\sigma_p(n)$ is the sum of the $p^{\rm th}$
powers of all divisors of $n$.  $E_4$ is a modular form of weight 
4 so $E_4(1/{T}) = T^4 E_4(T)$, while $E_2$ fails 
to be a modular form of weight two since
\beq
E_2\left(\frac1{T}\right) = -T^2 E_2(T) + \frac{6T}{\pi}\,.
\label{e2mod}
\eeq

Since there is already a constant term in the expansion of the
superpotential we will perform most of our analysis considering
$W_0=0$ which means no supersymmetry breaking by the fluxes
themselves. Contrary to the case with one single exponential, we will
find many minima without needing to tune the value of $W_0$ to obtain
large compactification volume naturally once we break
supersymmetry. We will see later how $W_0$ affects our results.

\subsubsection{Numerical analysis}\label{section5.2.1}

In order to perform reliable computations with the Eisenstein series
we use the `weak coupling' or large radius expansion of $W(T)$ in
eq.~(\ref{wdef}) only when $2\pi\, X > N$. For other ranges we can use
the property~(\ref{e2mod}). For instance, for $1 < 2\pi\, X < N$ we
can transform the $E_2({T}/{N})$ term to obtain
\beq
W(T)=E_2(T) + \frac{N}{T^2} E_2\left(\frac{N}{T}\right) - \frac{6}{\pi T}\,.
\label{w2def}
\eeq

\TABLE{\renewcommand{\tabcolsep}{2pt}
\begin{tabular}{|c|ccc|}\hline
$N$ & 2 & 3 & 4  
\\\hline
$X$  &  1.46 & 1.70 & 1.82 
\\\hline
$V_{\rm min}$ & $-1.73\cdot 10^{-2}$ & $-1.60\cdot 10^{-2}$  & 
$-1.32\cdot 10^{-2}$
\\\hline
\end{tabular}
\caption{Minima of the scalar potential.\label{tab1}}}

\noindent Similarly, when $2\pi\, X < 1$\footnote{We have to keep in mind that
  in these regimes there may be substantial corrections to the
  K\"ahler potential which are not under control.} we can transform
both terms in $W(T)$ to obtain
\beq
W(T)=-\frac{1}{T^2}E_2(T) + \frac{N}{T^2} E_2\left(\frac{N}{T}\right).
\label{w1def}
\eeq 

For $N \leq 4$ we have that the minima appear when $2\pi\, X > N$, so
that one has to use the expression~(\ref{w8def}) for the
superpotential. One finds the supersymmetric minima are at
$Y=N/2$ and $X$ (when $N\leq4$) at the values given in table~\ref{tab1}.

For $N \geq 5$ the flat direction at 
$Y={N}/{2}$ turns into a saddle point and pairs of 
supersymmetric minima $T_{i1}$, $T_{i2}$ on either side in the 
$Y$ direction appear, such that 
$Y_{i1} + Y_{i2} = N$ and $X_{i1} =X_{i2}$.
Examples of these are given in table~\ref{tab2}.

{\renewcommand\belowcaptionskip{-.5em}
\TABLE[t]{\begin{tabular}{|c|c|c|c|c|c|}
\hline
$N$ & 5 & 6 & 7 & 8 & 9 \\
\hline
$X$  & 1.48 & 1.65 & 1.77 & 1.85 & 1.88 \\
$Y$ & 2.24 & 2.40 & 2.63 & 2.88 & 3.13 \\
\hline
$V_{\rm min}$ & $-1.35 \cdot 10^{-2}$ & $-1.37\cdot 10^{-2}$
&$-1.27\cdot 10^{-2}$ & $-1.16\cdot 10^{-2}$ & $-1.07\cdot 10^{-2}$ \\
\hline
\end{tabular}%
\caption{Minima of the scalar potential for several cases.\label{tab2}}} 
}

For $N \geq 10$ we have that the minima begin to appear when $2\pi\, X
< N$, so that one has to use the expression~(\ref{w2def}) for the
superpotential to compute the scalar potential~(\ref{spot}). This
needs to be done in order to have a convergent series expansion
in~(\ref{efns}). Using this we find a lot of supersymmetric minima in
the scalar potential, noting that the number of minima increases with
$N$.  We list some of the results table~\ref{tab3}.

{\renewcommand\belowcaptionskip{-.5em}
\renewcommand\tabcolsep{5pt}
\TABLE[t]{\begin{tabular}{|ccc|ccc|ccc|}
\hline
& $N=30$& & &$N=40$& & &$N=50$& \\
\hline
X& Y & $V_{\rm min}$& X & Y & $V_{\rm min}$& X & Y & $V_{\rm min}$\\
\hline
1.20&6.44 & $-1.12\cdot 10^{-2} $&
1.28&6.95 & $-5.72\cdot 10^{-3}$&
1.10&7.52 & $-3.43\cdot 10^{-3}$
\\
1.32&11.71 & $-1.34\cdot 10^{-2} $& 
1.44&8.73 & $-1.68\cdot 10^{-2}$& 
1.15&14.36 & $-8.58\cdot 10^{-3}$
\\
1.57&8.40 & $-1.70\cdot 10^{-2} $& 
1.60&15.31 & $-1.54\cdot 10^{-2} $&
1.36&8.92 & $-1.36\cdot 10^{-2}$
\\
1.65&12.15 & $-1.16\cdot 10^{-2} $& 
1.76&11.32 & $-1.46\cdot 10^{-2} $&
1.59&10.98 & $-1.70\cdot 10^{-2}$ 
\\
2.08&5.69 & $-5.21\cdot 10^{-3} $& 
1.84&16.77 & $-1.06\cdot 10^{-2} $& 
1.76&18.97 & $-1.39\cdot 10^{-2}$
\\
 & & & 
2.17&6.56 & $-4.19\cdot 10^{-3} $& 
1.82&14.24 & $-1.26\cdot 10^{-2}$ 
\\
 & & &  & & & 1.92&21.30 & $-9.10\cdot 10^{-3}$
\\
 & & &  & & & 2.24&7.32 & $-3.51\cdot 10^{-3}$ 
\\\hline
\end{tabular}%
\caption{Minima of the scalar potential for $N=30,40,50$.\label{tab3}}}
}

{\renewcommand\belowcaptionskip{-1em}
\DOUBLEFIGURE[t]{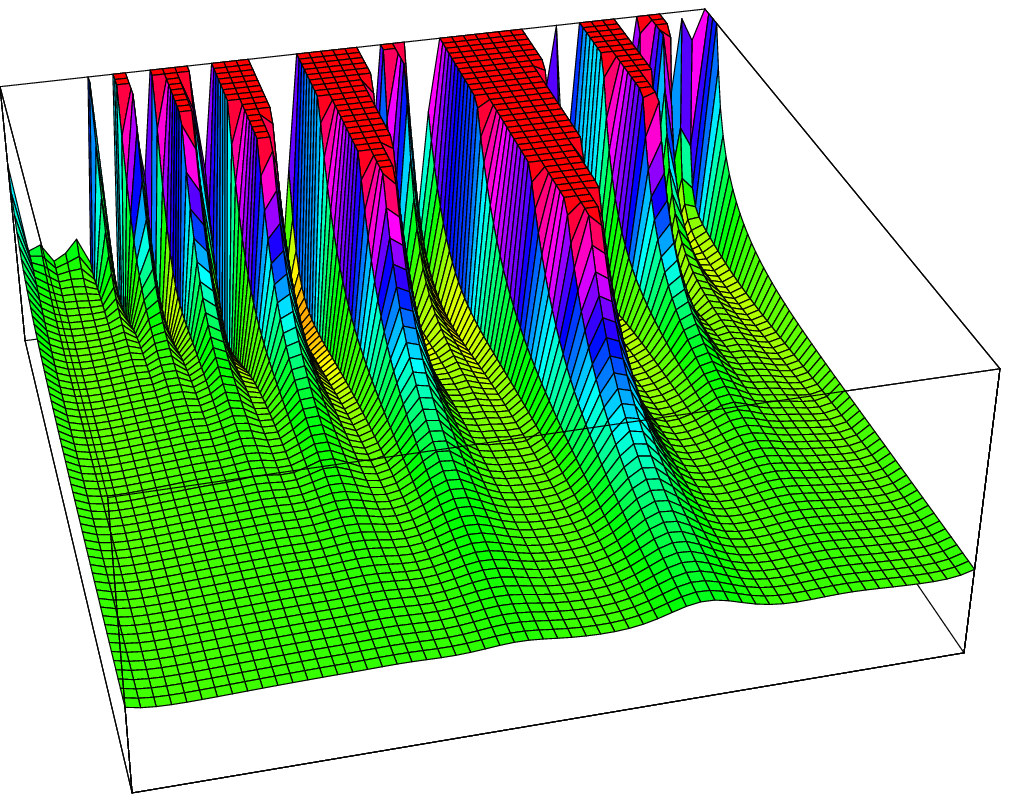,width=.48\textwidth}
{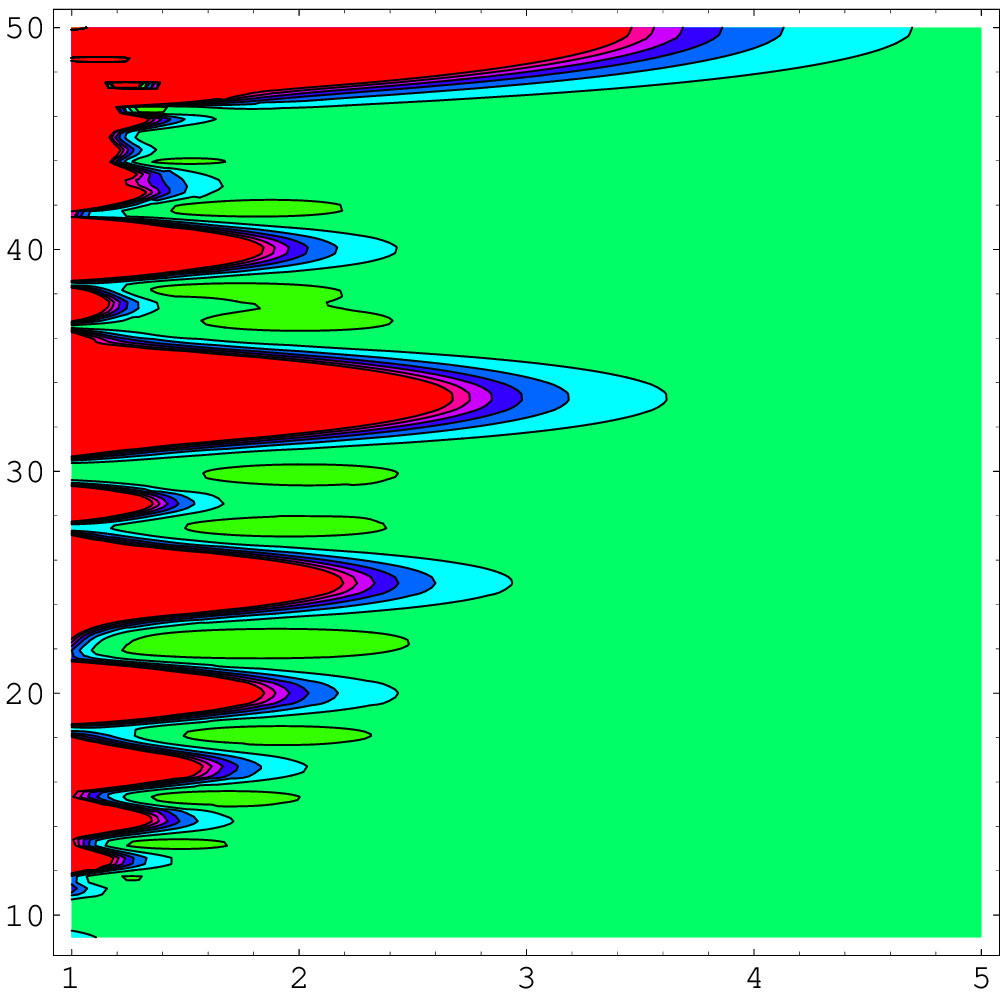, width=.48\textwidth}
{Graph for the scalar potential with $N=100$.\label{fig5}}
{Contour graph for the scalar potential with $N=100$.\label{fig6}}
}

\looseness=-1Also we present in figure~\ref{fig5} a 3D plot of the potential for
the case $N=100$ illustrating the rich structure of the potential with
many supersymmetric AdS minima. In figure~\ref{fig6} we present a
contour plot showing the location of the minima and how the landscape
changes in field space.

To illustrate the fact that we find an increasing number of minima
when we increase $N$, and also other interesting facts, we present in
table~\ref{tab4} some other results of the numerical analysis. Note that by
${\rm Min}_G$ in table~\ref{tab4} we denote the global minimum of the potential and
by ${\rm Min}_X$ we denote the minimum with a largest (finite) value of $X$.

{\renewcommand\belowcaptionskip{-1em}
\renewcommand\arraystretch{.96}
\TABLE[t]{\begin{tabular}{|c|c|ccc|ccc|ccc|}
\hline
$N$ &  Number  of &  & ${\rm Min}_G$ & $ $ &  & 
${\rm Min}_X$ & $ $\\
$ $ & ${\rm Minima}$ & X & Y & $V_{\rm min}$ & X & Y & $V_{\rm min}$
\\ \hline
10 & 2 &  1.28 & 3.59 & $-1.11 \cdot 10^{-2}$ & 1.85 & 3.32 & 
$-1.01 \cdot 10^{-2}$
\\ \hline
20 & 3 & 1.76 & 7.84 & $-1.43 \cdot 10^{-2}$ & 1.96 & 4.65 & 
$-6.92 \cdot 10^{-3}$
\\ \hline
30 &  5 & 1.57 & 8.40 & $-1.70 \cdot 10^{-2}$ & 2.08 & 5.69 & 
$-5.21 \cdot 10^{-3}$
\\ \hline
40 &  6 & 1.44 & 8.73 & $-1.69 \cdot 10^{-2}$ & 1.28 & 6.95 & 
$-5.73 \cdot 10^{-3}$
\\ \hline
50 &  7 & 1.59 & 10.98 & $-1.70 \cdot 10^{-2}$ & 1.10 & 7.52 & 
$-3.46 \cdot 10^{-3}$
\\ \hline
60 &  10 & 1.45 & 10.78 & $-1.70 \cdot 10^{-2}$& 1.20 & 7.62 & 
$-1.62 \cdot 10^{-3}$
\\ \hline
70 &  10 & 1.53 & 12.62 & $-1.76 \cdot 10^{-2}$& 2.32 & 8.61 & 
$-2.67 \cdot 10^{-3}$
\\ \hline
80 &  10 & 1.62 & 14.45 & $-1.69 \cdot 10^{-2}$ & 1.13 & 8.85 & 
$-1.07 \cdot 10^{-3}$
\\ \hline
90 & 10  & 1.49 & 13.75 & $-1.78 \cdot 10^{-2}$ & 2.38 & 9.73 & 
$-2.16 \cdot 10^{-3}$
\\ \hline
100 &  10 & 1.55 & 15.28 & $-1.76 \cdot 10^{-2}$ & 2.41 & 10.25 & 
$-1.97 \cdot 10^{-3}$
\\ \hline
200 &  15 & 1.52 & 20.98 & $-1.76 \cdot 10^{-2}$ & 2.56 & 14.37 & 
$-1.06 \cdot 10^{-3}$
\\ \hline
300 &  19 & 1.60 & 26.00 & $-1.74 \cdot 10^{-2}$ & 2.70 & 17.50 & 
$-7.26 \cdot 10^{-4}$
\\ \hline
400 & 20 & 1.49 & 29.59 & $-1.80 \cdot 10^{-2}$& 2.68 & 20.20 & 
$-5.53 \cdot 10^{-4}$
\\ \hline
500 & 25 & 1.47	& 32.20 & $-1.77 \cdot 10^{-2}$& 2.70 & 22.60 & 
$-4.46 \cdot 10^{-4}$
\\ \hline
600 & 25 & 1.50	& 36.31 & $-1.79 \cdot 10^{-2}$& 2.73 & 24.70 & 
$-3.74 \cdot 10^{-4}$
\\ \hline
700 & 27 & 1.47 & 37.80 & $-1.75 \cdot 10^{-2}$& 2.75 & 26.60 & 
$-3.22 \cdot 10^{-4}$
\\ \hline
800 & 31 & 1.62	& 45.70 & $-1.72 \cdot 10^{-2}$& 2.77 & 28.43 & 
$-2.83 \cdot 10^{-4}$
\\ \hline
900 & 34 & 1.56	& 46.10 & $-1.76 \cdot 10^{-2}$& 2.78 & 30.10 &
$-2.52 \cdot 10^{-4}$
\\ \hline
1000 & 38 & 1.48 & 46.50 & $-1.72 \cdot 10^{-2}$& 2.78 & 31.80 & 
$-2.28 \cdot 10^{-4}$
\\ \hline
\end{tabular}%
\caption{Number of minima for different values of N in the 
supersymmetric case.\label{tab4}}}
}

From the information shown in table~\ref{tab4} we may extract some
general remarks regarding the minima. First, the number of minima
grows quite fast when $N$ grows. Also the potential at the minimum
with larger value of $X$ increases with $N$. Furthermore, even though
we do not have a general proof, for all the cases we have analysed,
all the vacua of this theory happen to be supersymmetric.

Finally, although it is not necessary in our case,
following~\cite{kklt} we have studied the effect of turning on
$W_0$. Notice that with $W_0=0$ we have many minima but all of them
correspond to not too large values of the compactification scale $X$.
We have found that tuning $W_0$ has the effect of allowing minima with
very large values of $X$ improving the validity of the field
theoretical analysis of the potential. Otherwise the general structure
of the potential remains the same.

\vspace{-1pt minus 3pt}

\subsubsection{Analytical considerations}\label{section5.2.2}
\vspace{-1pt minus 2pt}

The scalar potential~(\ref{spot}) appears to be, in general terms, too
complicated to do an analytical study of its minima for a
superpotential such as the $\cn=1^*$ one. In fact, introducing the
superpotential~(\ref{w2def}) into the condition~(\ref{mnsloc}) that is
used to find the supersymmetric minima we arrive to non linear
equations that cannot be solved analytically.

\pagebreak[3]

In any case, some interesting features appear when we study the
behaviour of the potential in the limit $(2\pi\,X)\,N \gg X^2+Y^2$. In
that limit the superpotential $W$ can be well approximated by the
function
\beq
W \sim 1 + \frac{N}{T^2} - \frac{6}{\pi T}\,.
\label{waprox}
\eeq
This approximation is derived from~(\ref{w2def}) just by keeping the
constant terms in the $E_2$ expansions and neglecting the exponential
terms.

Using the approximate superpotential in~(\ref{waprox}), we find that
the condition for a supersymmetric minimum, i.e.,\ $2XW^\prime
- 3W = 0$, reduces to a cubic equation in $X$ which can be solved 
analytically. This cubic equation is given by:
\beq
\left(\frac{288}{\pi}+2N\pi\right) X-168 X^2+24\pi 
X^3=18 N\,,
\label{x}
\eeq
with  $Y$ given in terms of $X$ by the following relation:
\beq
\label{y}
Y = \sqrt{N+\frac{X(3\pi X-16)}{\pi}}\,.
\eeq

The full form of the solutions for $X$ and $Y$ is not very
enlightening, but they allow an expansion in powers of $1/N$ that
happens to be more useful.\footnote{The equation~(\ref{x}) has of
  course three solutions for $X$, but only one of the three solutions
  of the cubic equation gives a positive $X_{\rm min}$ in the range of
  validity of the approximation.}
\beqa
X_{\rm min}&=&\frac{9}{\pi}-\frac{3240}{\pi^3}\frac{1}{N}+\frac{5015520}
{\pi^5}\frac{1}{N^2}+\cdots
\\
Y_{\rm min}&=&\sqrt{N}\left(1+\frac{99}{2\pi^2}\frac{1}{N}-
\frac{502281}{8\pi^4}\frac{1}{N^2}+\cdots\right)
\eeqa

It is clear from these expressions that when $N$ is large the solution
for $X$ will tend asymptotically to $9/\pi$, and $Y$ will tend to
$\sqrt{N}$. It is also possible to compute the value of the potential
in this minimum (also as an expansion in power series of $N$)
substituting the approximate expression for the superpotential $W$
given in~(\ref{waprox}) into~(\ref{spot}).  Its expansion in powers of
$1/N$ is given by
\beq
\label{v}
V_{\rm min}=-\frac{2\pi}{27}\frac{1}{N}+\frac{50}{3\pi}\frac{1}{N^2}+\cdots\,.
\eeq
In fact, this minimum is the minimum with the largest finite value of
$X$. Also we have found that for $N\ge50$ these
results~(\ref{x}),~(\ref{y}) and~(\ref{v}) agree well with those
obtained in the numerical analysis (see table~\ref{tab4}). This
minimum is in fact the minimum that appears to have the largest value
of the potential for large values of $N$. This will play an important
role in the next section, where we explore the minima of the potential
when the supersymmetry is broken.

\subsection{Supersymmetry breaking}\label{section5.3}

In this subsection we will study the changes produced in the structure
of the vacua when we introduce in the scalar potential the
supersymmetry breaking term~(\ref{sb}). We will see that with the
introduction of such a term it is possible to lift the vacua from anti
de Sitter vacua to de Sitter vacua, for some range of the parameter
$D$.\footnote{Recall that the $\cn=1^*$ superpotential has a mass
  scale $m^3$. This was irrelevant for the discussion of the
  supersymmetric case since it could be rescaled out of the potential,
  affecting only its absolute value at the minima. We have to keep
  this in mind when considering the range of variation of the
  parameters $W_0$ and $D$ which will carry now an extra scale
  determined by $m$.}

\subsubsection{Analytical considerations}\label{section5.3.1}

As in the previous case, the scalar potential~(\ref{potproto}) appears
to be, in general terms, too complicated to do an analytical
description of its minima. In fact, introducing the
superpotential~(\ref{w2def}) into the scalar
potential~(\ref{potproto}) and minimizing it to find the minima leads
to non linear equations that cannot be solved analytically.

In any case, similarly to the previous cases, some interesting
features appear when one studies the behaviour of the potential in the
limit $2\pi\,X\,N\gg(X)^2+(Y)^2$. As in the previous case, in that
limit the superpotential $W$ can be well approximated by the
function~(\ref{waprox}).

Using the approximate superpotential in~(\ref{waprox}), we find that
the condition for minimum reduces to two polynomial equations in
$X\,,\,Y$. Those equations cannot be solved in general but for $N$
large it is easy to find that $Y$ is given in terms of $X$ by the
following relation:
\beq
\label{yy}
Y = \sqrt{\frac{5N\pi X+6 X^2 (\pi X-7)}{6(\pi X-1)}}\,.
\eeq

Using this relation~(\ref{yy}), we find a complicated polynomial
equation on just $X$. This equation can be simplified for large $N$ to
a cubic polynomial on $X^2$, and therefore can be solved
analytically. In fact, what we found is that only two of the solutions
are positive and then lead to two real positive solutions for
$X$. Nevertheless just one of these two solutions is a minimum, the
other one corresponding to a saddle point.  The solution found is
proportional to $\sqrt{N}$, with the proportionality constant given in
terms of the supersymmetry breaking parameter $D$. That is, the
solutions for the minimum are given by $X_{\rm min}=f(D)\,N^{1/2}$.

The exact form of the function $f(D)$ is a complicated expression, but
in any case it is possible to extract some useful features from it. In
fact, it is easy to see that $f(D)$ is an increasing function on $D$,
until it reaches an upper bound $D_M$. For values $D> D_M$ the values
of the function $f(D)$ are no longer real, this meaning that the
potential would have no minima. Therefore, in the limit of validity of
the analytical analysis, there is a bound for the supersymmetry
breaking parameter $D$ if we want the potential to have minima. This
bound is given by
\beq
D<\frac{1}{360}(27+7\sqrt{21})\equiv D_{M}\,.
\eeq 
Therefore, for values of the supersymmetry breaking parameter $D$
larger than this bound we will not find any minima of the scalar
potential.

When the bound in $D$ is saturated we find that
\beq
\label{xxx}
X_{\rm min}^M=\sqrt{\frac{5 }{12(-4+\sqrt{21})}}\,{N}^{1/2}\,,
\eeq
These values for $X$ and $D$ allow an expansion in powers of $1/N$ of
the potential in the minimum. We find that the value of the potential
is positive and is given by
\beq
\label{vv}
V_{\rm min}^M=0.0406\frac{1}{N^{3/2}}+0.0386\frac{1}{N^2}+\cdots
\eeq

Also from the form of $f(D)$ it is clear that one can fine-tune the
supersymmetry breaking parameter $D$ so that the value of the scalar
potential in the minimum $X_{\rm min}$ approaches arbitrarily to
zero. In fact one can compute the value of $D$ such that $V_{\rm
  min}=0$. The full expression is not very useful, but it allows an
expansion in power series in $1/N$, which is given by
\beq
D_0=\frac{3}{20}-\frac{9}{20\pi}\sqrt{\frac{3}{5N}}+\frac{99}{1600\pi^2}
\,\frac{1}{N}+\cdots\,,
\eeq
and also the values of $X_{\rm min}$ and $Y_{\rm min}$ are
\beqa
X_{\rm min}^0&=&\sqrt{\frac{5}{12}N}+\frac{17}{8\pi}-\frac{99}
{64\pi^2}\sqrt{\frac{3}{5N}}+\cdots
\\
Y_{\rm min}^0&=&\sqrt{\frac{5}{4}N}-\frac{1}{8\pi\sqrt{3}}-\frac{3463}
{192\pi^2}\sqrt{\frac{1}{5N}}+\cdots
\eeqa
where the value of $Y_{\rm min}^0$ is obtained from~(\ref{yy}).  Also
remember that $V_{\rm min}^0=0$.

In fact, we have found that for $N\ge50$ these results agree well with
those obtained in the numerical analysis, as we will show in the
following subsection.

\subsubsection{Numerical results}\label{section5.3.2}

In this subsection we present the results found in the numerical
analysis of the scalar potential in the non-supersymmetric case.  We
have computed numerically the minima of the potential for $D=D_0$ for
several values of $N$, such as $N=50$, $N=100$, $N=500$ and
$N=1000$. The results obtained from the analysis are shown in
table~\ref{tab5}.

\TABLE{\begin{tabular}{|cccccc|}
\hline
 & $N=50$& & & $N=100$ &  
\\ \hline
X& Y & $V_{\rm min}$& X & Y & $V_{\rm min}$ 
\\ \hline
2.63&14.24&$1.71 \cdot 10^{-3}$&2.95&22.24& $1.23 \cdot 10^{-3} $ \\
4.46&20.48&$3.36 \cdot 10^{-4}$  &3.77&37.26&$5.36 \cdot 10^{-4}$ \\
5.46&7.94&0  &4.62&44.20&$1.35 \cdot 10^{-4}$ \\
 & & &7.35&11.24&0 
\\ \hline
 & $N=500$& & & $N=1000$ &  
\\ \hline
X& Y & $V_{\rm min}$& X & Y & $V_{\rm min}$ 
\\ \hline
2.88 & 47.61 & $1.88 \cdot 10^{-3}$& 3.02& 68.96& $1.74 \cdot 10^{-3} $ \\
3.40 & 52.65 & $1.18 \cdot 10^{-3}$& 5.89& 192.46& $9.23 \cdot 10^{-5}$ \\
5.43 & 118.30& $1.31 \cdot 10^{-4}$& 5.89& 207.54& $9.23 \cdot 10^{-5}$ \\
5.46 & 131.72& $1.30 \cdot 10^{-4}$& 6.69& 240.77& $4.18 \cdot 10^{-5}$ \\
6.31 & 158.00 & $4.62 \cdot 10^{-5}$& 6.69& 259.23& $4.18 \cdot 10^{-5}$ \\
6.30 & 175.33 & $4.62 \cdot 10^{-5}$& 8.24& 321.20& $1.38 \cdot 10^{-5}$ \\
8.49 & 237.24 & $8.59 \cdot 10^{-6}$& 8.24& 345.46& $1.38 \cdot 10^{-5}$ \\
15.33 & 25.11& 0& 11.46& 482.05& $2.42 \cdot 10^{-6}$ \\
 & & & 21.34& 35.51& 0
\\ \hline
\end{tabular}%
\caption{Minima of the scalar potential for several non-supersymmetric
  cases.\label{tab5}}}

As a matter of comparison with the supersymmetric case, we show in
figure~\ref{fig7} the scalar potential for $N=100$ with supersymmetry
broken.

Finally we show in table~\ref{tab6} a similar information as the one
shown in table~\ref{tab4} in the last subsection, where we write the
number of minima varying with $N$. In table~\ref{tab6} we denote by
${\rm Min}_{X,Y}$ the minima with largest value of $X,Y$. We can
notice that the number of minima still increases with $N$ but slower
than in the supersymmetric case. An argument for this is that the
non-supersymmetric term $\delta V$ tends to smooth out the potential
as it moves the minima up, so some of the minima eventually
disappear. Another implication of this term is that the minima tend to
have larger value of the compactification scale, large enough to
create a hierarchy as compared to the string scale, as desired
phenomenologicaly. Furthermore, unlike the supersymmetric case, the
minima appear to be ordered in $X$, where by ordered we mean that the
value of the potential decreases as $X$ increases. The one with
largest compactification scale has a smallest cosmological constant.

\EPSFIGURE{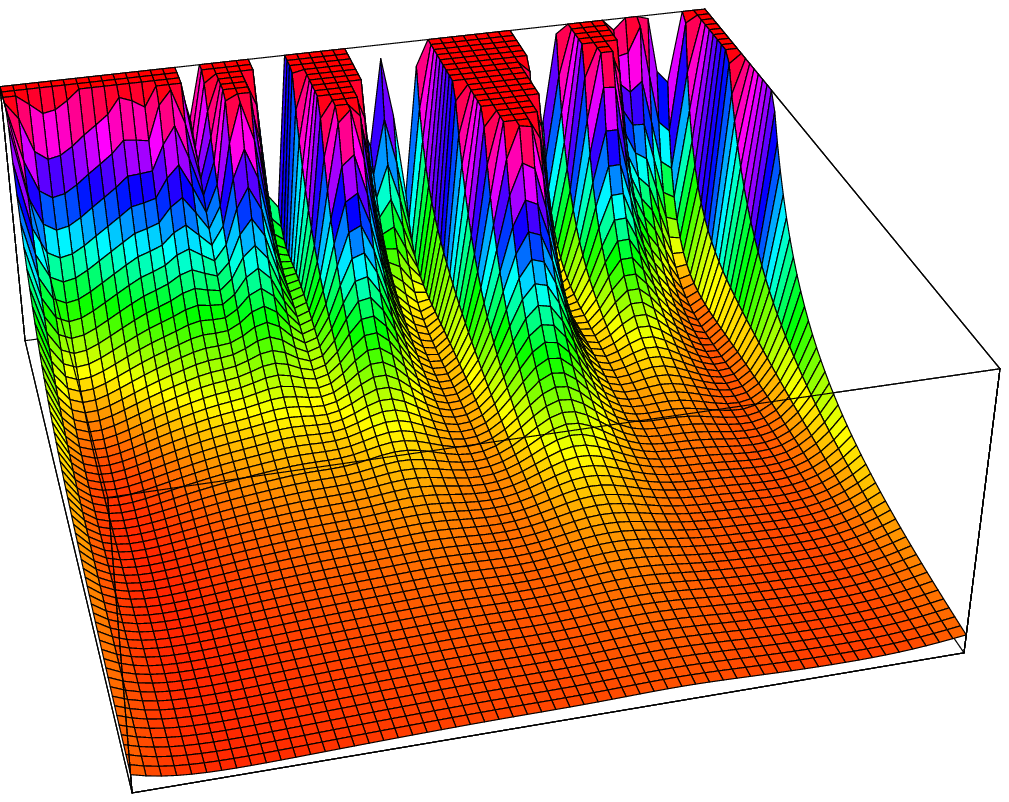, width=8cm}{Graph for the non-supersymmetric 
scalar potential with $N=100$.\label{fig7}}

{\sloppy In figure~\ref{fig8} we illustrate the effect of the
non-supersymmetric term in the potential. The value of the potential
at the minima is presented for several values of the
parameter~$D$. For $D=0$ we have the supersymmetric case with more
minima and not ordered, the increasing of the parameter $D$ will
reduce the number of minima, increase the value of the
compactification scale at the minima and order them from larger to
smaller values of the cosmological constant as the radius increases.

}

\TABLE{\begin{tabular}{|c|c|ccc|ccc|}
\hline
$N$ & ${\rm Number\  of}$ & $ $ & ${\rm Min}_X$& & $ $ &  ${\rm Min}_Y$ & \\
$ $ & ${\rm Minima}$ & X & Y & $V_{\rm min}$ & X & Y & $V_{\rm min}$
\\ \hline
10 & 2 & 3.98 & 3.90 & 0 & 3.98 & 3.90 & 0
\\ \hline
20 & 2 & 3.82 & 5.00 & 0 & 3.82 & 5.00 & 0
\\ \hline
30 &  2 & 4.42 & 6.10 & 0 & 2.88 & 12.00 & $7.69 \cdot 10^{-4}$
\\ \hline
40 & 2 & 4.99 & 7.10 & 0 & 3.49 & 16.30 & $4.60 \cdot 10^{-4}$
\\ \hline
50 &  3 & 5.46 & 7.94 & 0 & 4.46 & 20.48 & $3.36 \cdot 10^{-5}$
\\ \hline
60 & 3 & 5.89 & 8.70 & 0 & 4.06 & 25.50 & $3.09 \cdot 10^{-4}$ 
\\ \hline
70 &  4 & 6.29 & 9.40 & 0 & 4.17 & 30.10 & $2.42 \cdot 10^{-4}$
\\ \hline
80 &  4 & 6.63 & 10.00 & 0 & 4.28 & 34.80 & $1.92 \cdot 10^{-4}$
\\ \hline
90 & 4 & 7.04 & 10.70 & 0 & 4.55 & 39.40 & $1.60 \cdot 10^{-4}$
\\ \hline
100 & 4 & 7.35 & 11.24 & 0 & 4.62 & 44.20 & $1.35 \cdot 10^{-4}$
\\ \hline
200 &  5 & 10.02 & 16.00 & 0 & 5.94 & 91.80 & $4.30 \cdot 10^{-5}$
\\ \hline
300 & 5 & 12.03 & 19.40 & 0 & 6.61 & 107.50 & $1.13 \cdot 10^{-4}$ 
\\ \hline
400 & 6 & 13.76 & 22.40 & 0 & 7.73 & 124.50 & $7.52 \cdot 10^{-5}$
\\ \hline
500 & 6 & 15.40 & 25.10 & 0 & 8.54 & 156.70 & $5.48 \cdot 10^{-5}$ 
\\ \hline
600 & 8 & 16.70 & 27.50 & 0 & 9.29 & 211.00 & $4.21 \cdot 10^{-5}$
\\ \hline
700 & 8 & 17.97 & 29.70 & 0 & 10.09 & 245.30 & $8.03 \cdot 10^{-5}$
\\ \hline
800 & 9 & 19.20 & 32.80 & 0 & 10.81 & 253.80 & $2.79 \cdot 10^{-5}$
\\ \hline
900 & 9 & 20.30 & 33.70 & 0 &  11.10 & 432.90 & $2.35 \cdot 10^{-5}$
\\ \hline
1000 & 9 & 21.34 & 35.51 & 0 & 11.46& 482.05& $2.42 \cdot 10^{-6}$ 
\\ \hline
\end{tabular}%
\caption{Minima for different values of N in a non-supersymmetric case
with $D=D_0$.\label{tab6}}}

{\renewcommand\belowcaptionskip{-.8em}
\footnotesize
\psfrag{D=0 }{$D=0$}
\psfrag{D=0.06}{$D=0.06$}
\psfrag{D=0.08}{$D=0.08$}
\psfrag{D=0.1 }{$D=0.1$}
\psfrag{D=0.12}{$D=0.12$}
\psfrag{D=Do \(=0.139\) }{$D=D_0(=0.139)$}
\EPSFIGURE[t]{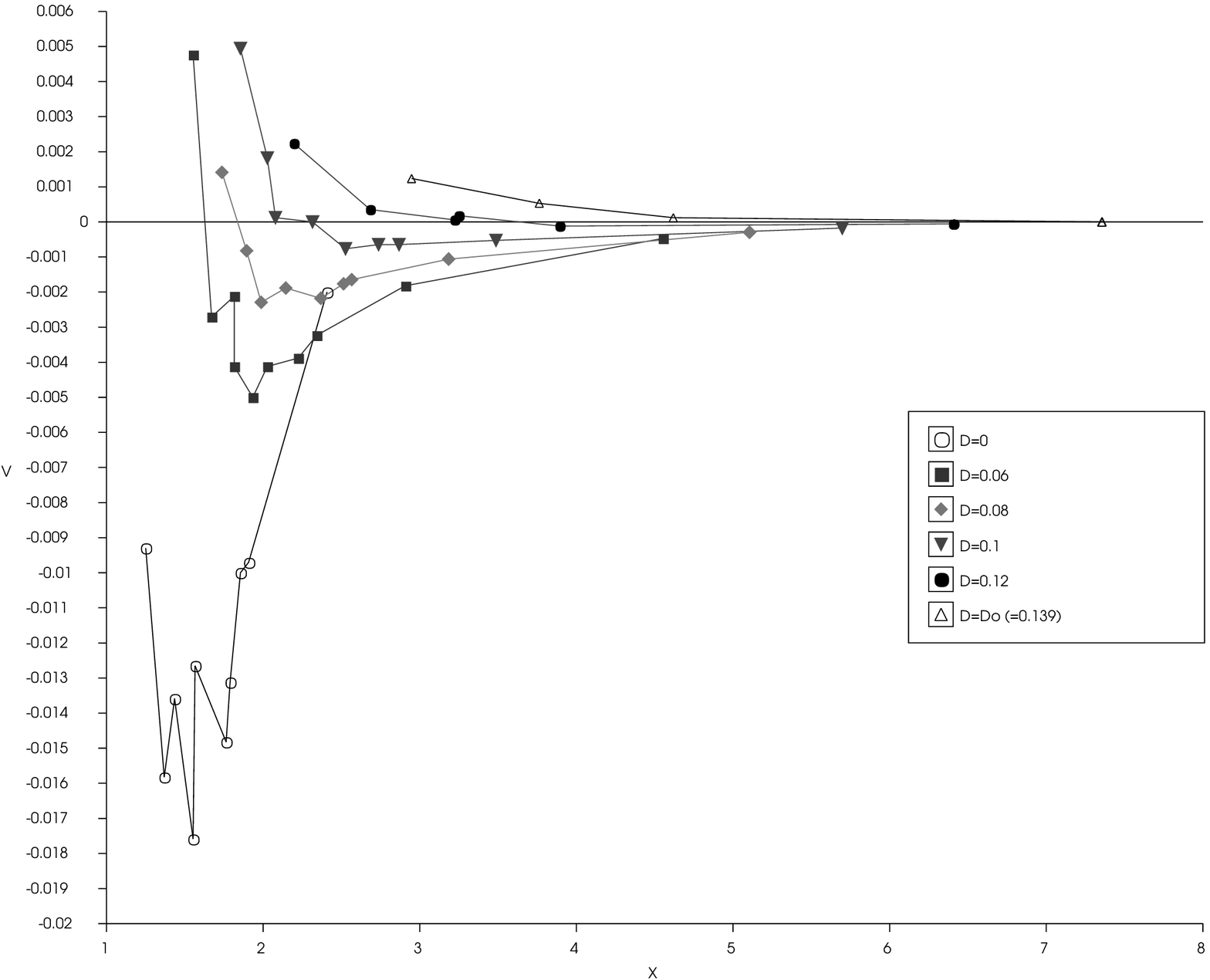,  width=\textwidth}{Potential versus $X$ for
$N=100$ and different values of $D$.\label{fig8}}
}

Turning on a nonvanishing value of $W_0$ tends to change some of these
conclusions.  The minimum with the largest value of $X$ can be lifted
to positive vacuum energy while some of the other ones remain anti de
Sitter. Besides this, the number of minima stays of the same order.

We can see that, in general, increasing the value of $N$ substantially
increases the number of vacua and the value of $X$ at the
minima. Notice that $N$ can be quite large in string theory and we may
attempt extracting results varying $N$ over many orders of
magnitude. However, there are limiations on the numerical analysis if
$N$ is too large, since many more terms in the series have to be
considered and the accuracy of the results would not be
guranteed. Also computer and woman time to perform the analysis would
increase substantially. Notice also that the number of minima we
present here is actually a lower bound since, again, the accuracy of
the numerical calculations could have missed some of them. In
particular, minima which would be very close to each other could be
missed.

\section{Stability of the vacua}\label{section6}

Having a situation with many nonsupersymmetric vacua with positive
cosmological constant implies that the relative stability of this
vacua should be considered. In particular if our own universe may be
described by one of them, it is important to know the lifetime of each
of the vacua against decay to any of the ones with smaller value of
the cosmological constant and compare it with the age of the universe.

In the general case with many vacua the stability question is very
much model and vacuum dependent. Given any particular minimum of the
potential there is a nonvanishing probability towards tunneling to any
of the other minima. The lifetime will depend on the height of the
potential and the side of the local maxima or saddle points that
separate two different minima. The general analysis is therefore very
complicated. We will limit here to the well controlled case of two
notrivial minima that we found in the case with two exponentials. This
may serve as an illustration of the techniques used in the general
case.

We know that the effective scalar potential has the standard runaway
behaviour towards infinity in the radial modulus $X$. The existence of
the runaway vacuum at $\infty$ in field space with zero energy is a
common feature of all string theories~\cite{ds}. This means that none
of the two minima that appear in the scalar potential (see
figures~\ref{fig3},~\ref{fig4}) will be stable, as they are unstable,
not only to decay from one minimum to the other minimum, but also to
decay to the runaway minimum at $X\rightarrow\infty$, by means of
either quantum tunneling or thermal excitation over a
barrier~\cite{cdl,hm}.

As we already mentioned, if we want to describe the present stage of
the acceleration of the universe using a model of the kind presented
here, we would like the last de Sitter minimum to be so that $V \sim
10^{-120}$ in Planck units. This can always be done in these models
just by adjusting the value of the different parameters of the
theory. In the case presented in section~\ref{section3}, we can
achieve this by fine-tuning $W_0$ and $D$, while in the $\cn=1^*$ such
a value for the cosmological constant can be obtained by slightly
modifying the value of $D=D_0$ (for which the value of the potential
is zero in the last minimum) by a small amount. Doing so one can
always get the desired value for $V$ in the last de Sitter minimum.
We will restrict here to the analysis of the case considered in
section~\ref{section3}, so we will have two minima with different
values of the scalar potential $V_0^{(+)}$, $V_0^{(-)}$ such that
$V_0^{(+)}>V_0^{(-)}\sim 10^{-120}$. Therefore we will have a model
with several decay possibilities. In fact from the minimum at
$V_0^{(+)}$ it is possible to decay either to the minimum at
$V_0^{(-)}$ or to the minimum at $X\rightarrow \infty$, while from the
minimum at $V_0^{(-)}$ it is just possible to decay to the minimum at
$X\rightarrow \infty$. We will also comment some features of the
$\cn=1^*$ case. In order to analyse this issue we will review several
features of tunneling theories taking into account gravitational
effects, following the original work of Coleman and De
Luccia~\cite{cdl}.

\subsection{Vacuum decay}\label{section6.1}

Let us consider a theory of a scalar field $\varphi$ with a potential
$V(\varphi)$ which has two local minima at $\varphi_{1},\,\varphi_{2}$
with $V(\varphi_1)>V(\varphi_2)$.  Both of the minima are stable
classically but the vacuum state at $\varphi=\varphi_1$ (that is, the
false vacuum) is unstable against quantum tunneling and will finally
decay into the true vacuum state at $\varphi=\varphi_2$, this vacuum
decay proceeding through the materialisation of a bubble of true
vacuum within the false vacuum phase. The tunneling action is given by
\beq
\label{accion}
S(\varphi)= \int d^4x
\sqrt{g}\left(-{1\over 2}R+{1\over 2} (\partial \varphi)^2 +
V(\varphi)\right),
\eeq
with a tunneling probability between two vacua, 
per unit time and unit volume, given by
\begin{equation}
\label{prob}
P(\varphi) \approx e^{-B}=e^{-S(\varphi)+S(\varphi_1)}\,.
\end{equation}
Here $\varphi$ is a solution of the equations of motion, which is
usually referred to as ``the bounce'', and $S(\varphi_1)$ is the
euclidean action in the initial configuration $\varphi=\varphi_1$.  In
the limit of small energy density between the two vacua Coleman and De
Luccia showed that the coefficient $B$ can be calculated in a closed
form. Although in the general case it is usually very difficult to
find analytical solutions for the Coleman-De Luccia instantons and
calculate the probability of decay through quantum tunneling, the
computation can be simplified within the range of validity of the
thin-wall approximation.  This approximation is valid when the
thickness of the transition region between the true and the false
vacuum is small compared with the radius of the bubble. In Minkowski
space, the condition $V_{\rm min}\ll V_{\rm max}$ usually means that
the thin-wall approximation is applicable, so the false vacuum state
will decay through the materialisation of a bubble of true vacuum
within the false vacuum phase, which is a quantum tunneling effect (by
$V_{\rm max}$ we denote the high of the de Sitter maximum that
separates two minima). That condition is usually fulfilled in the
model considered here, and we will explicitly check later that in
those cases we are always within the limits of the thin-wall
approximation.

Now let us consider the case in which we have a vacuum with positive
cosmological constant (a dS vacuum) and a vacuum with null
cosmological constant (a Minkowski vacuum).\footnote{This is always
  our case unless in the decay from $V_0^{(+)}$ to $V_0^{(-)}$, that
  are both de Sitter vacua. Nevertheless, as $V_0^{(+)} \gg
  V_0^{(-)}\sim10^{-120}$, it is a good approximation to consider that
  the decay is also from dS to Minkowski.} For those cases Coleman and
De Luccia found that the decay will be produced through a nucleation
of a bubble of radius\footnote{In units of $M_P=1$.}
$\bar{\rho}=\frac{12 S_1}{4\epsilon+3 S_1^2}$, where $\epsilon$
denotes the energy density difference between the two vacua $\epsilon=
V(\varphi_1)-V(\varphi_2)$. As the final vacuum has null cosmological
constant $V(\varphi_2)=0$, then $\epsilon=V(\varphi_1)$. Also $S_1$
denotes the tension of the bubble wall, that is given by
\begin{equation}\label{s1}
S_1 = \int_{\varphi_1}^{\varphi_2} d\varphi\, \sqrt{2 V(\varphi)}\,,
\end{equation}
The coefficient $B$ is then given by
\beq\label{coef}
B=\frac{24\pi/\epsilon}{(1+4\epsilon/3S_1^2)^2}\,.
\eeq
The thin-wall approximation is justified in this context if 
$\bar{\rho}$ and $\sqrt{3/\epsilon}$ are large compared with 
the range of variation of the scalar field $\varphi$, that is, 
$\bar{\rho},\sqrt{3/\epsilon} \gg \Delta\varphi$.

\subsection{Comparing tunneling probabilities}\label{section6.2}

In this subsection we will compute the probabilities of all possible
tunneling trajectories between the different vacua of the
potential~(\ref{potproto}),~(\ref{potdosexp}). Nevertheless before
doing so we need to make a remark: In our case we have a potential
that depends on two scalar fields. When two or more fields are
involved obtaining the bounce $\varphi$, and therefore computing
tunneling probabilities, becomes a much more difficult task (see for
example~\cite{kusenko}), except in trivial cases.

An example of a trivial case is to consider the tunneling from any of
the minima to the runaway minimum at $X\rightarrow \infty$. In those
cases the lines $Y=0,\pi/a$ are always minima of the potential in the
$Y$ direction, so we can consider a one-dimensional potential
$V^{(\pm,\infty)}=V(X,Y=0,\pi/a)$ and analyse the tunneling
probability in the standard way (we would obtain one-dimensional
graphs like the ones shown in figure~\ref{fig4}$b$). Also note that
the scalar field $\varphi$ is defined in such a way that the kinetic
term in the action~(\ref{accion}) is canonical. In our analysis, the
complex scalar field $T$ has a kinetic term that is given by the
derivatives of the K\"ahler potential~(\ref{kpdef}), and is given by
$\frac{3}{4(X)^2}(\partial X \partial X+ \partial Y \partial Y$). If
we consider that $Y$ is fixed then the kinetic term coming from the
K\"ahler potential reads $\frac{3}{4}(\partial\,\ln\, X)^2$, so the
canonical field would be of the form
$\varphi^{(\pm,\infty)}=\sqrt{\frac{3}{2}}\ln\, X$.

On the other hand, the case of computing the tunneling probability
from the minimum $V_0^{(+)}$ to the minimum $V_0^{(-)}$ is more
subtle, as we cannot trivially reduce it to the one-dimensional case
in the same way as the previous case. It is however possible to find a
lower bound for the tunneling probability by replacing the multi-field
potential by a suitably chosen single field potential, as any chosen
trajectory will have larger tension~(\ref{s1}) than the minimum
one. As we are interested in comparing probabilities such a bound will
be enough for us. Therefore we can consider that the line that joins
the two minima together is a good approximation for the bounce (as we
already mentioned, this is an upper bound for the tension, as the
bounce is the one that minimises the action).  Then we can consider
that $Y=\frac{\Delta Y^{(+,-)}}{\Delta X^{(+,-)}}X+b$ where by $\Delta
X,Y^{(+,-)}$ we denote the difference in the value of $X,Y$ between
both minima, clearly $\Delta Y^{(+,-)}={\pi}/{a}$, while the value
of $\Delta X^{(+,-)}$ cannot be written analytically.  If we use this
relation between the fields $X$ and $Y$ we recover again the standard
one-dimensional case where the canonical field will now be
$\varphi^{(+,-)}=\sqrt{\frac32(1+(\frac{\Delta Y^{(+,-)}}{\Delta
    X^{(+,-)}})^2)}\,\ln X$.

In order for this model to be useful for explaining the actual
acceleration stage of the universe and also the smallness of the
cosmological constant we should check that the tunneling probability
for the decay from $V_0^{(+)}$ to $V_0^{(-)}$ is larger than the
tunneling probability for the decay from $V_0^{(+)}$ to the minimum at
$V=0$, and also that the minimum at $V_0^{(-)}$ has a decay time
larger than the life of the universe.

Let us begin by computing the decay probability from the minimum at
$V_0^{(+)}$ to the minimum at $V=0$ and comparing it with the decay
probability from the minimum at $V_0^{(+)}$ to the minimum at
$V_0^{(-)}$. For those cases we will have from~(\ref{coef}) that the
probability will be given in both cases by
\beq
\label{p1}
P\approx \exp\left(-\frac{24\pi^2/ V_0^{(+)}}
{(1+4V_0^{(+)}/3S_1^2)^2}\right), 
\eeq
as $V_0^{(+)} \gg V_0^{(-)}$. Therefore it is clear that the decay with
smaller tension $S_1$ will be the most probable. The tension of the
bubble wall in both cases can be written as 
\beq
\label{tension2}
S_1^{(+,\infty)}\sim \sqrt{V_1^{(+,\infty)}}\,
\Delta\varphi^{(+,\infty)}\,,
\qquad
S_1^{(+,-)}\lesssim \sqrt{V_1^{(+,-)}}\,\Delta\varphi^{(+,-)}\,.
\eeq
where $V_1^{(+,\infty)}$, $V_1^{(+,-)}$ denote the height of the
maxima that separates any two minima (see figure~\ref{fig4}$a$
and~\ref{fig4}$b$).  Also $\Delta\varphi^{(+,\infty)}$,
$\Delta\varphi^{(+,-)}$ denote the variation of the canonical field in
each case. As can be shown in figure~\ref{fig4} we have that
$1\gtrsim\Delta\varphi^{(+,\infty)} \gg \Delta\varphi^{(+,-)}$ and
$V_1^{(+,\infty)} \gtrsim V_1^{(+,-)}$, so then we will have in
general terms that $S_1^{(+,\infty)}>S_1^{(+,-)}$. Therefore we find
that in this case it is more probable to decay to the minimum
$V_0^{(-)}\sim 10^{-120}$ than to the minimum $V=0$ at
$X\rightarrow\infty$.

Now we must compute the probability of decay from the minimum at
$V_0^{(-)}$ to the minimum at $V=0$. In this case the tension of the
bubble wall can be written as
\begin{equation}
\label{tension1}
S_1^{(-,\infty)} \sim \sqrt{V_1^{(-,\infty)}}\Delta\varphi^{(-,\infty)}\,,
\end{equation}
where $V_1^{(-,\infty)}$ is the height of the maximum that separates
the two minima. From the form of the potential shown in
figure~\ref{fig4}$a$, we can assume that $\Delta\varphi\sim {\cal
  O}(1)$, so it is clear from~(\ref{tension1}) that $S_1^2 \gg
V_0^{(-)}$. Therefore for the decay probability one simply gets
\begin{equation}
\label{p2}
P_{(-,\infty)}\approx \exp\left(-\frac{24\pi^2/ V_0^{(-)}}
{(1+4V_0^{(-)}/3S_1^2)^2}\right) 
\sim \exp\left(-\frac{24\pi^2}{V_0^{(-)}}+{64\pi^2\over S_1^2}+
\cdots\right).
\end{equation}
For $V_0^{(-)}\sim 10^{-120}$ this probability is extremely small, so
therefore, this dS vacuum can be considered stable in practical
terms. Also note that the thin-wall approximation is always true
within this model as ${\cal O}(1)\gtrsim \Delta\varphi$ and
$\bar{\rho}, (V_0^{(\pm)})^{-1/2} \gg 1$.

It is interesting to note that the analysis of the stability of the
last minimum in the $\cn=1^*$ is very similar to this last case. The
reason for that is the following: the analytical discussion of the
scalar potential in the non-supersymmetric case developed in
\pagebreak[3]section~\ref{section4} showed that apart from the minimum we also find
a saddle point located in (for $D=D_0$)
\beqa
\label{xx}
X_{sp}^0&=&1.12\sqrt{N}+1.32+1.33\frac{1}{\sqrt{N}}+\cdots
\\
Y_{sp}^0&=&1.44\sqrt{N}+0.37-1.36\frac{1}{\sqrt{N}}+\cdots
\eeqa
with a value of the potential given by
\beq\label{vvv}
V_{sp}^0=0.11\frac{1}{N^{3/2}}-0.46\frac{1}{N^2}+1.57\frac{1}{N^{5/2}}
+\cdots
\eeq
This value of the potential it is small for large $N$, but is still
big compared with $10^{-120}$ for reasonable values of $N$. As this
point is a maximum in one direction but a minimum in the other
directions, we can perform an analysis of the stability of the vacua
following the lines of the two exponential case. As in this case we
also find $\Delta\varphi \sim {\cal O}(1)$ and $V_{sp} \gg V_{\rm
  min}^0\sim 10^{-120}$ we will arrive to the same conclusion as the
one in~(\ref{p2}).

In the $\cn=1^*$ potentials, as in all the cases with many minima, a
typical situation may be that some of the minima would correspond to
de Sitter space and others to anti de Sitter. We may imagine living in
the one corresponding to de Sitter space with the smallest value of
the cosmological constant and would wonder about its decay probability
towards a global minimum with negative cosmological constant. As
discussed in~\cite{cdl}, the decay to a state of negative vacuum
energy may or may not occurr and may lead to gravitational collapse.
This has been recently reanalysed in~\cite{banks}.

Finally we would like to mention that it is also necessary to check
that the decay times of the de Sitter vacua are also not too long. The
fact that a de Sitter space has finite entropy introduces a time scale
that is the Poincar\'e recurrence time $t_r$~\cite{recurrence}. This
quantity is given by $t_r\sim e^{S_{dS}}$, where $S_{dS}$ denotes the
entropy of the de Sitter space. For dS space the entropy has a simple
sign-reversal relation with respect to the euclidean action calculated
for the false vacuum dS solution $\varphi=\varphi_{0}$, which is given
by $-{24\pi^2/ V_0}$. Then the recurrence time can be written as
$t_r\sim e^{24\pi^2/ V_0}$. An interesting property of this kind
of models is that the decay time of the de Sitter vacua never exceeds
the recurrence time of the de Sitter space $t_r$. This was first
noticed in~\cite{kklt} and can be easily checked from the following
expression
\begin{equation}
\frac{\ln t_r}{\ln t_{\rm decay}} = \ln {(1+
4V_0^{(\pm)}/3S_1^2)^2}>0\quad\Longrightarrow\quad t_r>t_{\rm decay}\,.
\end{equation}
The problems related to the decay time $t_{\rm decay}$ exceeding the recurrence 
time $t_r$ will then not appear in these models.

\section{Discussion}\label{section7}

We have presented examples of multiple de Sitter vacua in string
theory. Even though the examples we consider are still relatively
simple, they illustrate what can be expected from the general vacuum
structure of string theory, i.e.\ a multitude of vacua with different
values of the cosmological constant.

The parameters of the theory allow for one of the minima to have a
cosmological constant as small as we want. This requires fine
tuning\footnote{The fact that we do not need to have a nonvanishing
  superpotential from the fluxes ($W_0$) indicates we do not need to
  fine tune this quantity as in~\cite{kklt}. However we still need to
  fine tune the supersymmetry breaking parameter $D$ which even though
  is discrete it can be varied almost continuosly~\cite{kklt,frey}.}
but it can be ameliorated given the large number of minima, indicating
an anthropic approach to the cosmological constant problem, as
advocated by different authors~\cite{abbott,bp}. The number of minima
increases with the rank of the gauge group $N$. Furthermore since
there is an underlying $\SL(2,\IZ)$ symmetry behind the $\cn =1^*$
theory, we may expect that there could be further, possibly infinite,
minima if we explore other fundamental domains of this group. We have
essentially only explored $N$ copies of the strips defined by $-1/2<
Y<1/2$ for $X$ outside the unit circle in the upper half plane. For
each of these strips, the modular group has an infinite number of
fundamental domains that could indicate a huge multiplication of the
number of minima, inequivalent from the ones we found since $\SL(2,
\IZ)$ is not a symmetry of the theory.\footnote{Notice also that in
  computing the scalar potential we restricted ourselves to one single
  phase of the $\cn=1^*$ theory. In general we could consider any
  other values of $p$ and $q$.}  However their study is beyond the
limit of validity of our effective actions which are trusted only for
$X$ greater than the string scale. Furthermore our potentials have
periodicity $N$ in the $Y$ direction. Any correction that would
slightly break this periodicity could give rise to an infinite number
of minima.

The large number of vacua that can appear in these theories due to the
nontrivial superpotential complements the already rich structure of
vacua due to the presence of the fluxes~\cite{douglas}. The large
number of 3-cycles in typical Calabi-Yau manifolds imply a large
amount of possibilities. Remember also that although the combination
of the fluxes $KM$ is restricted by the tadpole cancellation
condition, the ration $K/M$ is a free (quantised)
parameter.\footnote{Notice that typical four-folds can have Euler
  number between $10^3$ and $10^6$~\cite{philip}, allowing for many
  different combinations of $M$ and $K$ to satisfy the tadpole
  condition.} This combination appears in the warp factor and allows
the tuning of the parameter $D$ to get a small cosmological constant,
defining the `discretuum' of vacua as described
in~\cite{susskind,bp,frey}. All these effects were present in the
single exponential case considered in KKLT. The large number of vacua
we found has to be multiplied by this degeneracy.  Although it may not
be large enough degeneracy for a naturally small cosmological
constant, the greater the number of minima the more natural is to find
a cosmological constant of the right size.

The fact that our potentials depend nontrivially on at least two real
fields $X, Y$ makes the discussion of the system more interesting than
for single field potentials in several ways. Since we may have many de
Sitter minima, if we imagine the universe starting in any of them, it
would leave naturally to different periods of inflation, either from
tunneling between minima but also by naturally rolling after the
tunnelings. There are so many valleys and hills in the potential that
it may not be impossible to find regions of slow roll between
different minima.

This combination of tunneling plus rolling has been considered in the
past on different models of inflation such as open
inflation~\cite{martin, lythbook}.  A detailed study of the
possibility for these potentials to give rise to realistic (eternal)
inflation would be clearly of great interest. For this we recall that
the main obstacle to have successfull inflation from string theory is
precisely the lack of control of the moduli potentials. In particular
the different proposals of D-brane inflation~\cite{dbraneinflation}
assume that there is an unknown stringy mechanism that fixes the
moduli and then, after that, D-brane inflation could occur. This is
clearly a very strong assumption and so far attempts of combining
D-brane inflation with moduli fixing have been running into
problems.\footnote{S.~Kachru and J.~Maldacena private
  communication~\cite{kachruetal}.} Therefore we may consider
seriously the possibility that actually the modulus $T$ could be the
inflaton field, and these potentials could give rise to interesting
combinations of inflationary processes. This is clearly a possible
subject for future investigation.

\looseness=1Another open question left unanswered here is the detailed analysis of
the decays of the different minima. This is complicated by the facts
that we have two-field potentials and that the fields do not have
canonically normalised kinetic terms. Our discussion was mostly
carried for the simplest case of two exponentials with only two
minima, but clearly a complete analysis for the $\cn=1^*$ and other
more general potentials would be needed.

Even though our models are relatively simple, they illustrate the
potential richness of the string landscape once the different `moduli'
fields acquire nontrivial potentials. We do not pretend that the
superpotentials discussed here, such as the one from the $\cn=1^*$
theory, would be particularly special over other more realistic
realisations of non-perturbative potentials. Actually, we regard it as
an interesting tool which includes all the ingredients expected from
non-perturbative physics, in particular the infinite instanton sum can
be under control thanks to the mathematical properties of modular
forms.  Other non-perturbative superpotentials recently derived,
including different deformations of $\cn=4$ theory~\cite{kumar2} would
be interesting to study.

The mechanism of KKLT, although very interesting, takes the simplest
class of models in which only one modulus is left unfixed by the
fluxes. In more realistic models we would expect many moduli left
unfixed by the fluxes and finding a many fields non-perturbative
potential for them would be an important challenge. We regard our
study here as a nontrivial, yet manageable, realisation of the
multi-vacua scenario in string theory and hope it may be of interest
to explore further properties of the space of string vacua.

\section{Note added in revised version}\label{section8}

The analysis presented in this article was performed using a
supersymmetry breaking term induced in the potential due to the
inclusion of an anti D3-brane in the configuration. Such a term was
taken to be given by $\delta V=D/X^3$, with $D$ a positive constant
depending on the fluxes turned on in the compactification.
Nevertheless, when this work was already finished it was pointed out
in~\cite{kachruetal} that this induced term is not the one to be used
if the anti D3 brane is in the warped region, as it is.  The correct
term should be $\delta V=\tilde{D}/X^2$ (with ${\tilde D}$ also a
positive, flux-dependent, constant).  This arises because the
supersymmetry breaking term coming from the inclusion of an anti
D3-brane in the warped compactifications considered here scales like
${\rm e}^{4A}/X^3$, and in the highly warped regime ${\rm e}^{4A}\sim
X{\rm e}^{-8\pi K/3g_sM}$.  This fact does not change the main
conclusions of the paper although it changes the numerical
analysis. We briefly summarize in this added note the changes produced
in the numerical analysis.

As it was found in the $D/X^3$ case, with the introduction of the
supersymmetry breaking term $\tilde{D}/X^2$ it is also possible to
lift the vacua from anti de Sitter vacua to de Sitter vacua, for some
range of the parameter $\tilde{D}$. Again we found that the effect of
the supersymmetry breaking term depends on the range of values of the
parameter ${\tilde D}$. If ${\tilde D}$ is very small (compared with
the value of the potential in the supersymmetric case), the potential
will not change substantially and the minima remain anti de
Sitter. For a critical value of ${\tilde D}$ one of the minima will
move up to zero vacuum energy and then to de Sitter space. Continuing
increasing ${\tilde D}$, more minima become de Sitter until all of
them are either de Sitter or Minkowski. We will denote by ${\tilde
  D}_0$ that precise value of the parameter ${\tilde D}$ such that all
the values of the potential at the minima are positive or zero. If
${\tilde D}>{\tilde D}_0$ the nonsupersymmetric term starts dominating
the potential and starts eliminating the different extrema to make the
potential runaway with $X$.

Actually, we have computed numerically the minima of the potential for
${\tilde D}={\tilde D}_0$ for several values of $N$, such as $N=50$,
$N=100$ and $N=500$.  The results obtained from the analysis are shown
in table~\ref{tab7}, where in those cases ${\tilde D}_0=0.04836,\
0.04958,$ and $0.04969$ for $N=50,\ 100,$ and $500$ respectively.

\TABLE[t]{\renewcommand\tabcolsep{5pt}\begin{tabular}{|ccccccccc|}
\hline
 & $N=50$& & & $N=100$ & & &$N=100$ & 
\\ \hline
X& Y & $V_{\rm min}$& X & Y & $V_{\rm min}$ & X & Y & $V_{\rm min}$
\\ \hline
1.70 & 10.98 & $1.28 \cdot 10^{-3}$ & 1.09 & 45.37 & $3.79 \cdot 10^{-2}$
&1.94 & 18.12 & 0\\
1.97 & 18.96 & $4.72 \cdot 10^{-4}$ & 1.49 & 13.20 & $9.10 \cdot 10^{-3}$
&2.06 & 27.47 & $7.14 \cdot 10^{-4}$\\
2.15 & 14.25 & 0 & 1.66 & 15.28 & $2.13 \cdot 10^{-3}$
&2.35 & 22.29 & $6.26 \cdot 10^{-4}$\\
2.65 & 21.30 & $9.33 \cdot 10^{-4}$ & 1.80 & 38.15 & $2.91 \cdot 10^{-3}$
&2.55 & 29.86 & $1.02 \cdot 10^{-3}$\\
 & & &1.80 & 38.15 & $2.91 \cdot 10^{-3}$ & 2.70 & 36.87 & 
$1.09 \cdot 10^{-3}$\\
 & & &1.82 & 41.84 & $9.05 \cdot 10^{-4}$& & & 
\\ \hline
& & & & $N=500$ &  & & & 
\\ \hline
X& Y & $V_{\rm min}$& X & Y & $V_{\rm min}$& X & Y & $V_{\rm min}$ 
\\ \hline
1.46 & 28.5 & $1.26 \cdot 10^{-2}$ & 2.13 & 108.6 & $4.01 \cdot 10^{-4}$
& 1.5 & 30.2 & $8.04 \cdot 10^{-3}$\\ 
2.20 & 113.7 & $3.84 \cdot 10^{-4}$ & 1.57 & 32.2 & $4.34 \cdot 10^{-3}$ 
& 2.23 & 47.6 & $3.62 \cdot 10^{-4}$\\
1.63 & 116.9 & $3.46 \cdot 10^{-3}$ & 2.23 & 59.7 & $1.12 \cdot 10^{-3}$ & 
1.64 & 132.1 & $7.10 \cdot 10^{-3}$\\
 2.25 & 65.4 & $7.58 \cdot 10^{-4}$ & 1.66 & 34.4 & $2.14 \cdot 10^{-3}$ 
& 2.48 & 68.1 & $9.44 \cdot 10^{-4}$\\
1.77 & 37.0 & $5.37 \cdot 10^{-4}$ & 2.51 & 52.7 & $9.66 \cdot 10^{-4}$ & 
1.78 & 138.4 & $2.14 \cdot 10^{-3}$\\
2.53 & 146.2 & $1.11 \cdot 10^{-3}$ & 1.79 & 89.0 & $1.97 \cdot 10^{-3}$ & 
2.58 & 74.8 & $9.41 \cdot 10^{-4}$\\
1.82 & 92.8 & $1.14 \cdot 10^{-3}$ & 2.60 & 139.5 & $1.09 \cdot 10^{-3}$ & 
1.85 & 134.4 & $5.09 \cdot 10^{-4}$\\
2.68 & 139.4 & $1.102 \cdot 10^{-3}$ & 1.90 & 40.0 & 0 & 2.91 & 87.3 & 
$1.15 \cdot 10^{-3}$\\
2.04 & 58.0 & $1.24 \cdot 10^{-3}$ & 2.95 & 79.3 & $1.11 \cdot 10^{-3}$ & 
2.05 & 43.5 & $8.48 \cdot 10^{-6}$
\\ \hline
\end{tabular}%
\caption{Minima of the scalar potential for several 
non-supersymmetric cases.\label{tab7}}}

\EPSFIGURE[r]{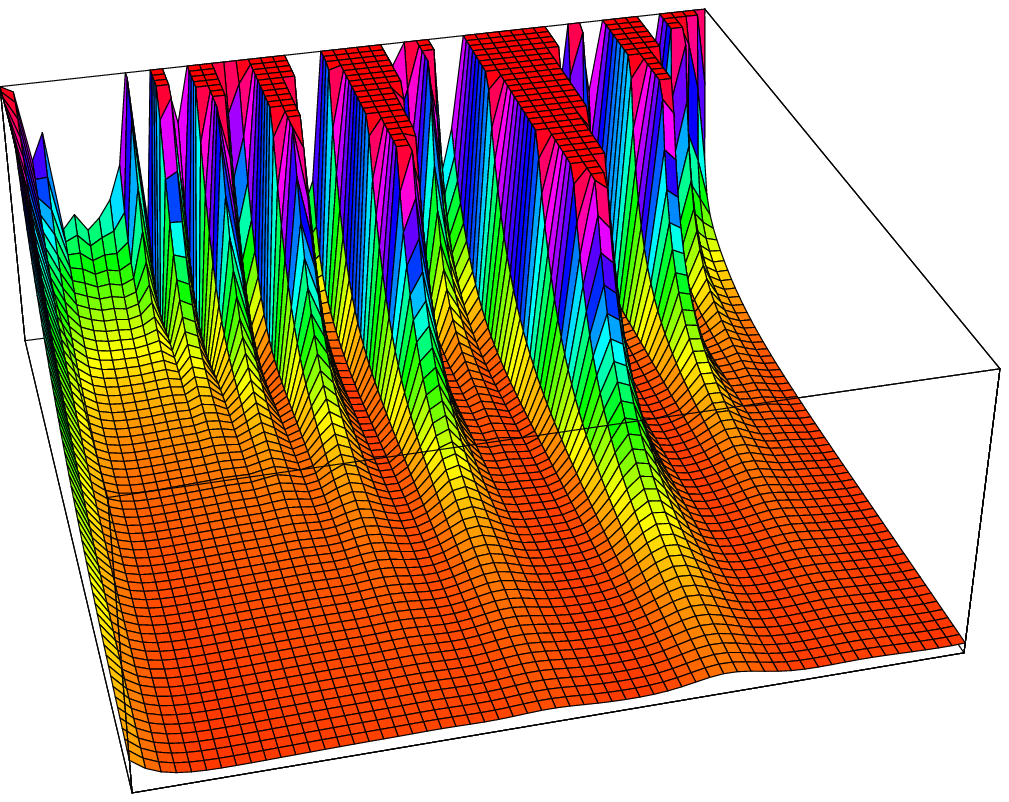,width=8cm}{Graph for the non-supersymmetric 
scalar potential with $N=100$.\label{fig9}}

Also we show in figure~\ref{fig9} the scalar potential for $N=100$
when the supersymmetry is broken. Note that comparing with
figure~\ref{fig7} the main differences are that now the potential is
less smooth and therefore has more minima (this can also be seen by
comparing the information shown in table~\ref{tab5} and
table~\ref{tab7}).

\TABLE[b]{\renewcommand\tabcolsep{5pt}
\begin{tabular}{|c|c|ccc|ccc|cc|}
\hline
$N$ & ${\rm Number}$ &  & ${\rm Min}_X$& &  &  
${\rm Min}_Y$ &  &  ${\rm Min}_{V=0}$ & 
\\
$ $ & ${\rm Minima}$ & X & Y & $V_{\rm min}$ & X & Y & $V_{\rm min}$
& X & Y
\\ \hline
10 & 1 & 2.4 & 3.31 & 0 & 2.4 & 3.31 & 0 & 2.4 & 3.31
\\ \hline
20 & 2 & 3.41 & 4.83 & $1.07 \cdot 10^{-3}$ & 3.41 & 4.83 & 
$1.02 \cdot 10^{-3}$ & 1.95 & 7.84
\\ \hline
30 &  2 & 2.33 & 12.13 & 0 &  2.33 & 12.13 & 0 & 2.33 & 12.13
\\ \hline
40 & 4 & 2.54 & 16.74 & $1.01 \cdot 10^{-3}$ & 2.54 & 16.71 &
 $1.01 \cdot 10^{-3}$ & 1.94 & 11.32
\\ \hline
50 &  4 & 2.65 & 21.32 & $9.35 \cdot 10^{-4}$ & 2.65 & 21.31 & 
$9.35 \cdot 10^{-4}$ & 2.15 & 14.25
\\ \hline
60 & 5 & 2.91 & 25.88 & $9.74 \cdot 10^{-4}$ & 2.91 & 25.88 & 
$9.74 \cdot 10^{-4}$ & 1.84 & 13.23
\\ \hline
70 &  6 & 3.26 & 30.52 & $1.14 \cdot 10^{-3}$ & 3.26 & 30.52 &
 $1.14 \cdot 10^{-3}$ & 1.98 & 15.53
\\ \hline
80 &  8 & 3.62 & 35.09 & $1.10 \cdot 10^{-3}$ & 3.62 & 35.09 &
 $1.10 \cdot 10^{-3}$ & 2.10 & 17.74
\\ \hline
90 & 8 & 2.61 & 33.38 & $8.76 \cdot 10^{-4}$ & 1.73 & 37.73 &
 $1.59 \cdot 10^{-3}$ & 1.82 & 16.28
\\ \hline
100 & 10 & 2.7 & 36.87 & $1.10 \cdot 10^{-3}$ & 1.82 & 41.84 & 
$9.06 \cdot 10^{-4}$ & 1.94 & 18.12
\\ \hline
200 &  21 & 2.84 & 53.81 & $1.13 \cdot 10^{-3}$ & 2.84 & 53.85 & 
$1.13 \cdot 10^{-3}$ & 1.88 & 87.82
\\ \hline
300 & 23 & 2.84 & 123.86 & $1.20 \cdot 10^{-3}$ & 2.26 & 124.10 & 
$1.38 \cdot 10^{-3}$ & 1.94 & 31.54
\\ \hline
400 &  22 & 3.15 & 75.72 & $1.15 \cdot 10^{-3}$ & 2.54 & 83.10 & 
$9.56 \cdot 10^{-4}$ & 1.88 & 87.71
\\ \hline
500 &  27 & 2.84 & 123.81 & $1.10 \cdot 10^{-3}$ & 2.26 & 124.10 & 
$1.38 \cdot 10^{-3}$ & 1.94 & 31.50
\\ \hline
600 & 28 & 3.34 & 95.55 & $1.19 \cdot 10^{-3}$ & 2.81 & 167.72 & 
$1.19 \cdot 10^{-3}$ & 1.92 & 44.41
\\ \hline
700 & 29 & 3.43 & 76.41 & $1.54 \cdot 10^{-3}$ & 2.27 & 148.15 & 
$5.10 \cdot 10^{-4}$ & 2.05 & 55.16
\\ \hline
800 & 29 & 3.43 & 76.41  & $1.54 \cdot 10^{-3}$  & 3.21 & 109.92 & 
$1.18 \cdot 10^{-3}$ & 2.05 & 55.16
\\ \hline
900 & 29 & 3.46 & 133.21 & $1.14 \cdot 10^{-3}$ & 3.38 & 133.21 & 
$1.18 \cdot 10^{-3}$ & 2.04 & 58.05
\\ \hline
1000 & 32 & 3.15 & 120.73 & $1.19 \cdot 10^{-3}$ & 3.07 & 129.22 & 
$1.19 \cdot 10^{-3}$ & 2.02 & 60.59
\\ \hline
\end{tabular}%
\caption{Minima for different values of N in a non-supersymmetric 
case with ${\tilde D}={\tilde D}_0$.\label{tab8}}}

{\sloppy In table~\ref{tab8} we show a similar information as the one shown in
table~\ref{tab6}, where we write the number of minima varying with
$N$. In table~\ref{tab8} we denote by ${\rm Min}_{X,Y}$ the minima
with largest value of $X,Y$ and by ${\rm Min}_{V=0}$ the minima with
vanishing value of the potential. We can notice that the number of
minima increases with $N$ more or less in the same rate than in the
supersymmetric case and then faster than in the $D/X^3$ case. The
reason for this is that the value of ${\tilde D}$ for which all the
anti de Sitter minima of the potential get lifted is smaller than in
the $D/X^3$ case, and therefore in the present case the $\delta V$
term smooths less the potential.

}
{\footnotesize
\psfrag{D = 0}{$D=0$}
\psfrag{D = 0.015}{$D=0.015$}
\psfrag{D = 0.03}{$D=0.03$}
\psfrag{D = 0.04}{$D=0.04$}
\psfrag{D = Do = 0.049}{$D=D_0=0.049$}
\EPSFIGURE[t]{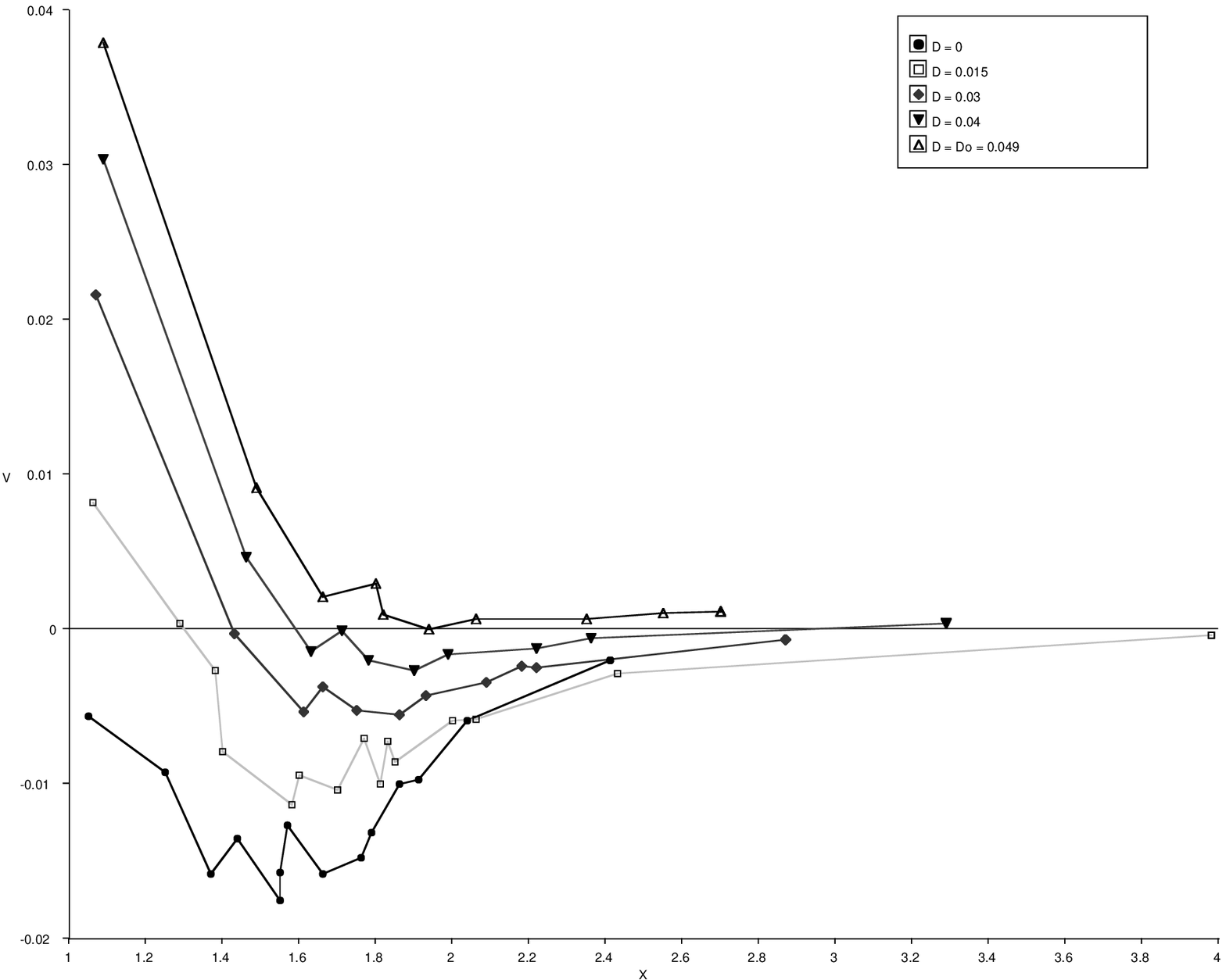, width=\textwidth}{Potential versus $X$ for
$N=100$ and different values of ${\tilde D}$.\label{fig10}}
}

Also, in figure~\ref{fig10} we illustrate the effect of the corrected
non-supersymmetric term in the potential. The value of the potential
at the minima is presented for several values of the parameter
${\tilde D}$.  For ${\tilde D}=0$ we have the supersymmetric case, the
increasing of the parameter ${\tilde D}$ will reduce the number of
minima and increase the value of the compactification scale at the
minima.  Note that now the minima are not ordered for increasing $X$,
and also that the values of the compactification scale at the minima
are, in general, smaller than in the $D/X^3$ case but still large
enough so that our approximations remain valid.

Summarising: the difference with the $1/X^3$ potentials is that now
there are more de Sitter minima, they are no longer ordered with
respect to the size of the extra dimension and the value of the volume
at the minima tends to be smaller than in the $1/X^3$ case, but still
large enough for the large radius approximation to be trustable.

Finally we would like to point out that the $1/X^3$ behaviour of the
nonsupersymmetric part of the potential presented in the main text is
still relevant. This is due to the recent proposal in~\cite{bkq} in
which instead of introducing an anti D3 brane on the throat, a flux of
magnetic fields on the D7 branes is considered. This adds an extra
term in the potential which happens to be proportional to $1/X^3$ if
the D7 brane is not sitting on the throat and to $1/X^2$ if the D7
brane is on the throat. In each case the rest of the analysis is as
presented in here and therefore both potentials are relevant depending
on the location of the D7 branes.

\acknowledgments

We would like to thank P.~Berglund, J.J.~Blanco-Pillado, M.~Bucher,
C.~Burgess, L.~Ib\'a\~nez, S.~Kachru, A.~Linde, R.~Rabad\'an and
A.~Uranga for useful conversations. The work of M.G.-R.  is supported
by Fundaci\'on Ram\'on Areces. C.E. is funded by EPSRC and F.Q.  by
PPARC.


\begin{thebibliography}{4}

\bibitem{fluxes}
A.~Strominger, \emph{Superstrings with torsion},
\npb{274}{1986}{253};\\
B.~de~Wit, D.J. Smit and N.D. Hari~Dass, \emph{Residual supersymmetry
  of compactified $D = 10$ supergravity}, \npb{283}{1987}{165};\\
K.~Becker and M.~Becker, \emph{M-theory on eight-manifolds},
\npb{477}{1996}{155} [\hepth{9605053}];\\
S.~Sethi, C.~Vafa and E.~Witten, \emph{Constraints on low-dimensional
  string compactifications}, \npb{480}{1996}{213} [\hepth{9606122}];\\
T.R. Taylor and C.~Vafa, \emph{RR flux on Calabi-Yau and partial
  supersymmetry breaking}, \plb{474}{2000}{130} [\hepth{9912152}];\\
B.R. Greene, K.~Schalm and G.~Shiu, \emph{Warped compactifications in
  M and F theory}, \npb{584}{2000}{480} [\hepth{0004103}];\\
C.S. Chan, P.L. Paul and H.~Verlinde, \emph{A note on warped string
  compactification}, \npb{581}{2000}{156} [\hepth{0003236}];\\
M.~Gra\~na and J.~Polchinski, \emph{Supersymmetric three-form flux
  perturbations on $AdS_5$}, \prd{63}{2001}{026001}
[\hepth{0009211}];\\
S.S. Gubser, \emph{Supersymmetry and F-theory realization of the
  deformed conifold with three-form flux}, \hepth{0010010};\\
A.R. Frey and J.~Polchinski, \emph{$N = 3$ warped compactifications},
\prd{65}{2002}{126009} [\hepth{0201029}];\\
P.K. Tripathy and S.P. Trivedi, \emph{Compactification with flux on K3
  and tori}, \jhep{03}{2003}{028} [\hepth{0301139}];\\
S.P. de~Alwis, \emph{On potentials from fluxes}, \hepth{0307084}.

\bibitem{drs}
K.~Dasgupta, G.~Rajesh and S.~Sethi, \emph{M-theory, orientifolds and
  G-flux}, \jhep{08}{1999}{023} [\hepth{9908088}].

\bibitem{gvw}
S.~Gukov, C.~Vafa and E.~Witten, \emph{CFT's from Calabi-Yau
  four-folds}, \npb{584}{2000}{69} [\hepth{9906070}].

\bibitem{gkp}
S.B. Giddings, S.~Kachru and J.~Polchinski, \emph{Hierarchies from
  fluxes in string compactifications}, \prd{66}{2002}{106006}
[\hepth{0105097}].

\bibitem{kst}
S.~Kachru, M.B. Schulz and S.~Trivedi, \emph{Moduli stabilization from
  fluxes in a simple IIB orientifold}, \jhep{10}{2003}{007}
[\hepth{0201028}].

\bibitem{kklt}
S.~Kachru, R.~Kallosh, A.~Linde and S.P. Trivedi, \emph{De~Sitter
  vacua in string theory}, \prd{68}{2003}{046005} [\hepth{0301240}].

\bibitem{dorey}
N.~Dorey, \emph{An elliptic superpotential for softly broken $N=4$
  supersymmetric Yang-Mills theory}, \jhep{07}{1999}{021}
[\hepth{9906011}].

\bibitem{ps}
J.~Polchinski and M.J. Strassler, \emph{The string dual of a confining
  four-dimensional gauge theory}, \hepth{0003136}.

\bibitem{susskind}
L.~Susskind, \emph{The anthropic landscape of string theory},
\hepth{0302219}.

\bibitem{abbott}
L.F. Abbott, \emph{A mechanism for reducing the value of the
  cosmological constant}, \plb{150}{1985}{427}.

\bibitem{tb}
J.D. Brown and C.~Teitelboim, \emph{Neutralization of the cosmological
  constant by membrane creation}, \npb{297}{1988}{787}.

\bibitem{bp}
R.~Bousso and J.~Polchinski, \emph{Quantization of four-form fluxes
  and dynamical neutralization of the cosmological constant},
\jhep{06}{2000}{006} [\hepth{0004134}].

\bibitem{salta}
J.L. Feng, J.~March-Russell, S.~Sethi and F.~Wilczek, \emph{Saltatory
  relaxation of the cosmological constant}, \npb{602}{2001}{307}
[\hepth{0005276}].

\bibitem{krasnikov}
N.V. Krasnikov, \emph{On supersymmetry breaking in superstring
  theories}, \plb{193}{1987}{37}.

\bibitem{noscale}
E.~Cremmer, S.~Ferrara, C.~Kounnas and D.V. Nanopoulos,
\emph{Naturally vanishing cosmological constant in $N=1$
  supergravity}, \plb{133}{1983}{61};\\
J.R. Ellis, A.B. Lahanas, D.V. Nanopoulos and K.~Tamvakis,
\emph{No-scale supersymmetric standard model}, \plb{134}{1984}{429}.

\bibitem{kpv}
S.~Kachru, J.~Pearson and H.~Verlinde, \emph{Brane/flux annihilation
  and the string dual of a non-supersymmetric field theory},
\jhep{06}{2002}{021} [\hepth{0112197}].

\bibitem{dv}
R.~Dijkgraaf and C.~Vafa, \emph{Matrix models, topological strings and
  supersymmetric gauge theories}, \npb{644}{2002}{3}
[\hepth{0206255}];
\emph{On geometry and matrix models}, \npb{644}{2002}{21}
     [\hepth{0207106}];
\emph{A perturbative window into non-perturbative physics},
\hepth{0208048}.

\bibitem{kumar2}
N.~Dorey, T.J. Hollowood, S.~Prem~Kumar and A.~Sinkovics, \emph{Exact
  superpotentials from matrix models}, \jhep{11}{2002}{039}
[\hepth{0209089}];
\emph{Massive vacua of $N=1^*$ theory and s-duality from matrix
  models}, \jhep{11}{2002}{040} [\hepth{0209099}].

\bibitem{fkq}
A.~Font, M.~Klein and F.~Quevedo, \emph{The dilaton potential from
  $N=1^*$}, \npb{605}{2001}{319} [\hepth{0101186}].

\bibitem{ds}
M.~Dine and N.~Seiberg, \emph{Is the superstring weakly coupled?},
\plb{162}{1985}{299}.

\bibitem{cdl}
S.R. Coleman and F.~De~Luccia, \emph{Gravitational effects on and of
  vacuum decay}, \prd{21}{1980}{3305}.

\bibitem{hm}
S.W. Hawking and I.G. Moss, \emph{Supercooled phase transitions in the
  very early universe}, \plb{110}{1982}{35}.

\bibitem{kusenko}
A.~Kusenko, \emph{Improved action method for analyzing tunneling in
  quantum field theory}, \plb{358}{1995}{51} [\hepph{9504418}].

\bibitem{banks}
T.~Banks, \emph{Heretics of the false vacuum: gravitational effects on
  and of vacuum decay, II}, \hepth{0211160}.

\bibitem{recurrence}
L.~Dyson, J.~Lindesay and L.~Susskind, \emph{Is there really a
  de~Sitter/CFT duality}, \jhep{08}{2002}{045} [\hepth{0202163}];\\
L.~Dyson, M.~Kleban and L.~Susskind, \emph{Disturbing implications of
  a cosmological constant}, \jhep{10}{2002}{011} [\hepth{0208013}];\\
N.~Goheer, M.~Kleban and L.~Susskind, \emph{The trouble with de~Sitter
  space}, \jhep{07}{2003}{056} [\hepth{0212209}].

\bibitem{douglas}
S.~Ashok and M.R. Douglas, \emph{Counting flux vacua},
\hepth{0307049}.

\bibitem{frey}
A.R. Frey, M.~Lippert and B.~Williams, \emph{The fall of stringy
  de~Sitter}, \prd{68}{2003}{046008} [\hepth{0305018}].

\bibitem{philip}
P.~Candelas, E.~Perevalov and G.~Rajesh, \emph{Toric geometry and
  enhanced gauge symmetry of F-theory/heterotic vacua},
\npb{507}{1997}{445} [\hepth{9704097}];\\
A.~Klemm, B.~Lian, S.S. Roan and S.T. Yau, \emph{Calabi-Yau fourfolds
  for M- and F-theory compactifications}, \npb{518}{1998}{515}
[\hepth{9701023}].

\bibitem{martin}
M.~Bucher, A.S. Goldhaber and N.~Turok, \emph{An open universe from
  inflation}, \prd{52}{1995}{3314} [\hepph{9411206}].

\bibitem{lythbook}
A.~Liddle and D. Lyth, \emph{Cosmological inflation and large scale
  structure}, Cambridge University Press, Cambridge 2000.

\bibitem{dbraneinflation}
For a review with many references see F.~Quevedo, \emph{Lectures on
  string/brane cosmology}, \cqg{19}{2002}{5721} [\hepth{0210292}].

\bibitem{kachruetal}
S.~Kachru, R.~Kallosh, A.~Linde, J.~Maldacena, L.~McAllister and
S.~P.~Trivedi, \emph{Towards inflation in string theory},
\newjournal{JCAP}{JCAPA}{0310}{2003}{013} [\hepth{0308055}].

\bibitem{bkq}
C.P. Burgess, R.~Kallosh and F.~Quevedo, \emph{De~Sitter string vacua
  from supersymmetric d-terms}, \jhep{10}{2003}{056}
[\hepth{0309187}].

\end{thebibliography}
\end{document}